\documentclass[aps,prd,twocolumn,superscriptaddress,nofootinbib,floatfix]{revtex4-2}
\usepackage{amsmath,amssymb,amsfonts}
\usepackage{graphicx}

\begin{document}

\title{Minimal-length kinematics of coherent light\\
and a proposed two-color vacuum chronoscopy}

\author{Zhi-Wei Wang}
\email{zhiweiwang.phy@gmail.com}
\affiliation{Department of Physics and Astronomy, University of Lethbridge, 4401 University Drive, Lethbridge, Alberta T1K 3M4, Canada}
\affiliation{College of Physics, Jilin University, Changchun 130012, People's Republic of China}
\affiliation{Computer Science, University of York, York YO10 5GH, United Kingdom}

\author{Saurya Das}
\email{saurya.das@uleth.ca}
\affiliation{Department of Physics and Astronomy, University of Lethbridge, 4401 University Drive, Lethbridge, Alberta T1K 3M4, Canada}
\affiliation{Quantum Horizons Alberta, University of Lethbridge, 4401 University Drive, Lethbridge, Alberta T1K 3M4, Canada}

\author{Samuel L.\ Braunstein}
\email{sam.braunstein@york.ac.uk}
\affiliation{Computer Science, University of York, York YO10 5GH, United Kingdom}

\begin{abstract}
The unification of quantum theory and general relativity suggests the
existence of a minimal measurable length, leading to a generalized
uncertainty principle (GUP) and an energy-dependent vacuum group
velocity. Testing these effects with macroscopic objects typically
encounters the ``soccer-ball problem,'' where applying Planck-scale
deformations to the total momentum predicts observationally ruled-out
velocity deviations. We adopt the prescription, natural in quantum field
theory (QFT), in which the GUP modifies single-mode commutators, and verify by explicit calculation
that the macroscopic group velocity of a multimode coherent pulse
reduces to the single-photon GUP velocity evaluated at
the central wavelength, multiplied by the classical Gaussian-beam
factor, with no dependence on the total pulse energy. The
energy--momentum relation $E=pc$ itself is unmodified; the deformation
enters through the velocity--momentum relation and is equivalent to a
modified dispersion relation in the wave vector, with a quadratic
coefficient that depends on which of the two variables is used. The
soccer-ball problem $\,$---$\,$ which arises only when the GUP is
applied to the center-of-mass momentum of a composite system $\,$---$\,$
is therefore absent for coherent light, and macroscopic petawatt pulses
probe the same parameter space as single high-energy photons. Building
on this framework, we propose a two-color vacuum chronoscopy experiment
using a short-pulse petawatt laser and its high-harmonic generation. By
systematically varying the macroscopic beam waist, our setup
separates Planck-scale kinematic shifts from classical
diffraction delays, establishing a controlled,
source-independent terrestrial methodology whose sensitivity we assess
against current astrophysical bounds and against the dominant
waist-independent systematics that limit it, residual-gas dispersion and the harmonic birth-time offset.
\end{abstract}

\maketitle

\section{Introduction}
\label{sec:intro}

The unification of quantum theory and general relativity is a
long-standing open problem in physics. Although there are several
candidate quantum theories of gravity, the sparsity of testable
predictions limits the development of these theories. One of the
important and apparently ubiquitous predictions of quantum theories of
gravity is the existence of a minimal measurable
length~\cite{gross1988,amati1989,maggiore1993,garay1995,kempf1995,amelino2002}.
Indeed, such a minimal length is also required by black hole physics,
which is widely accepted as one of the key pillars underlying the
development of quantum gravity~\cite{scardigli1999}. Incorporating the
effect of this minimal length into Heisenberg's uncertainty principle
yields the so-called generalized uncertainty principle
(GUP)~\cite{gross1988,amati1989,maggiore1993,garay1995,kempf1995,amelino2002}.
Since the uncertainty principle is fundamental in our current
understanding of quantum physics, the phenomenological impact of generalizing such a
principle is correspondingly broad~\cite{hossenfelder2013}.

Consequently, a phenomenological program has emerged to constrain the GUP parameters across a vast range of energy and length scales. In the low-energy regime, GUP effects have been shown to induce measurable shifts in well-understood quantum systems, allowing high-precision tabletop experiments---such as those measuring the Lamb shift, Landau levels, and currents in scanning tunneling microscopes---to place competitive bounds on the minimal length~\cite{das2011}. Simultaneously, advances in macroscopic quantum manipulation have inspired proposals using massive optomechanical oscillators to probe Planck-scale commutators directly in the laboratory~\cite{pikovski2012}. On cosmological fronts, Planck-scale deformations of the Bekenstein-Hawking entropy and Friedmann equations have been investigated as a possible mechanism for generating the baryon asymmetry of the early Universe~\cite{das2022baryon}. More recently, theoretical mechanisms tracking the temporal evolution and free expansion of macroscopic quantum wavepackets of general distributions have been proposed to probe these Planck-scale modifications dynamically~\cite{das2023wavepackets}.

Among these diverse signatures, a direct and widely discussed consequence of the GUP is the modification of the fundamental energy-momentum dispersion relation, predicting an energy-dependent vacuum group velocity for free particles~\cite{amelino1998}. In the formulation adopted here the relation $E=pc$ is retained and the deformation appears in the velocity--momentum relation; the two descriptions coincide at leading order. While this effect is actively constrained in astrophysics using high-energy gamma rays accumulated over cosmological distances---such as observations of distant gamma-ray bursts by the Fermi-LAT telescope~\cite{abdo2009, vasileiou2013}---terrestrial laboratory tests remain desirable for their repeatability, strict control over initial states, and avoidance of complex astrophysical emission models.

The interpretation of GUP-type modifications on macroscopic scales is delicate. Naively applying the modified commutator to the total center-of-mass momentum $P_{\mathrm{tot}} = \sum_i p_i$ of a composite system predicts velocity deviations that scale extensively with the number of constituents and are observationally ruled out. This is the so-called ``soccer-ball problem''~\cite{hossenfelder2014}. The natural quantum field theory (QFT) prescription, in which the deformation acts at the level of individual mode commutators, avoids this problem by construction: each constituent carries the same Planck-scale correction, and macroscopic dynamics inherits the single-particle dispersion. In this paper we show by explicit calculation that this expectation is realized for coherent light. Modeling a high-intensity laser pulse as a multimode coherent state~\cite{fedorov2017}, we evaluate the GUP-modified group velocity expectation and find that it reduces to the classical pulse velocity computed with the GUP-modified single-photon velocity--momentum relation. The macroscopic energy enters numerator and denominator homogeneously and drops out, leaving a velocity deviation set by the single-photon wavelength $\lambda$. This provides a paradox-free framework in which terrestrial petawatt-laser experiments probe the same Planck-scale parameter space as single high-energy astrophysical photons. Detailed derivations, the comparison with astrophysical bounds, and the budget of waist-independent systematics are collected in the Appendices.

\section{Generalized uncertainty principle}
\label{sec:gup}

String theory, loop quantum gravity, and black hole physics generally suggest a minimal measurable length $l_\text{min}$ of the order of the Planck length $l_\text{Pl} = \sqrt{\hbar G/c^3} \approx 1.6 \times 10^{-35}$~m. This can be phenomenologically implemented by modifying the canonical commutation relation between the physical position $x$ and momentum $p$ to~\cite{adv}
\begin{equation}
[x,p] = i\hbar \left( 1 - \alpha_0 \frac{ l_\text{Pl} }{\hbar} \, p + \beta_0 \Bigl(\frac{ l_\text{Pl}}{\hbar}\Bigr)^{\!2}  p^2 - \cdots \right),
\label{eq:GUP}
\end{equation}
where $\alpha_0$ and $\beta_0$ parameterize the linear and quadratic Planck-scale geometric corrections, respectively. We take Eq.~(\ref{eq:GUP}) as a generic phenomenological parameterization of linear-plus-quadratic minimal-length deformations~\cite{kempf1995,amelino2002,adv}, with $\alpha_0$ and $\beta_0$ treated as independent parameters. Their normalization is fixed by Eq.~(\ref{eq:GUP}), and therefore refers to an expansion in the \emph{physical} momentum $p$; this distinction matters at quadratic order and is made explicit below. In the specific three-dimensional algebra of Ref.~\cite{adv} the two parameters are not independent: the one-dimensional reduction along the propagation axis fixes $\alpha_0 = 2\alpha_0^{\rm ADV}$ and $\beta_0 = \alpha_0^2$. Bounds quoted here apply to the generic parameterization and specialize to any particular model by matching conventions.

Through the Hamiltonian equations of motion, $v_g = \frac{1}{i\hbar} [x, H(p)]$, this modified commutator leads to a momentum-dependent group velocity. For a highly relativistic particle ($E \approx p c$), the modified group velocity is \cite{das2011,hossenfelder2014}
\begin{equation}
v_g \approx \frac{ p c^2}{E} \left( 1 - \alpha_0 \frac{ l_\text{Pl} }{\hbar} \, p + \beta_0 \Bigl(\frac{ l_\text{Pl}}{\hbar}\Bigr)^{\!2}  p^2 - \cdots \right).
\label{eq:GUP_vg_single}
\end{equation}

Because photons admit no position operator in the Newton--Wigner sense, we do not rely on the commutator route for light. The deformation can instead be stated entirely in momentum space, with no position operator entering. Write the physical momentum as a map $p(p_0)$ of the translation generator $p_0 = \hbar k$, which photons possess irrespective of their localization properties. Since $[x,p] = i\hbar\, dp/dp_0$, the choice
\begin{equation*}
p = p_0 \Bigl[ 1 - \tfrac{1}{2}\alpha_0 \tfrac{ l_\text{Pl} }{\hbar} p_0
+ \bigl(\tfrac{1}{3}\beta_0 + \tfrac{1}{6}\alpha_0^2\bigr) \Bigl(\tfrac{ l_\text{Pl}}{\hbar}\Bigr)^{\!2} p_0^2 - \cdots \Bigr]
\end{equation*}
reproduces Eq.~(\ref{eq:GUP}) exactly, order by order in $l_\text{Pl}$, for arbitrary $\alpha_0$ and $\beta_0$. The energy $E(k) = c\,p(k)$ is then a deformed function of the wave vector, and the group velocity $d\omega/dk = c\,dp/dp_0$ is well defined for a wave:
\begin{equation*}
v_g = c \Bigl( 1 - \alpha_0 l_\text{Pl} k + \beta_0^{(k)} l_\text{Pl}^2 k^2 - \cdots \Bigr),
\qquad \beta_0^{(k)} \equiv \beta_0 + \tfrac{1}{2}\alpha_0^2 .
\end{equation*}
This is Eq.~(\ref{eq:GUP_vg_single}) rewritten in the wave-vector variable, and the rewriting is not innocuous at second order. The linear coefficient is unchanged, but $\beta_0$ governs the velocity as a function of photon energy $E = cp$, which is the variable in which astrophysical time-of-flight limits are quoted, whereas $\beta_0^{(k)}$ governs it as a function of wavelength $\lambda = 2\pi/k$, which is the variable the experiment proposed below controls. The two coincide whenever $\alpha_0 = 0$. Equation~(\ref{eq:GUP_vg_single}) is therefore equivalent to a modified dispersion relation of the type introduced in Ref.~\cite{amelino1998}, and the derivation that follows should be read as a consistency check of the per-mode prescription for macroscopic coherent states, not as a new single-photon prediction.

For a macroscopic system, directly substituting the total macroscopic momentum $P_{\text{tot}} = \sum p_i$ into Eq.~(\ref{eq:GUP_vg_single}) leads to the soccer-ball problem, predicting a deviation in the propagation speed that is ruled out by observation.

\section{Group velocity of multimode coherent light pulses}
\label{sec:pulse}

Within the QFT-natural prescription, the GUP modifies the commutator of each mode independently, so the soccer-ball problem does not arise. To compute the macroscopic group velocity of a high-intensity pulse, we evaluate the modified dispersion as an expectation value over the single-photon momenta in a multimode coherent state~\cite{fedorov2017}
\begin{equation}
\vert{\Psi}\rangle = \prod_{\vec{k},\sigma} e^{-\frac{|\alpha_{\vec{k},\sigma}|^2}{2}} \sum_{n_{\vec{k},\sigma}} \frac{(\alpha_{\vec{k},\sigma} a_{\vec{k},\sigma}^\dagger)^{n_{\vec{k},\sigma}}}{n_{\vec{k},\sigma} ! } \vert{0}\rangle,
\end{equation}
where $a_{\vec{k},\sigma}^\dagger$ creates a photon with wavevector $\vec{k}$ and polarization $\sigma$, and $\alpha_{\vec{k},\sigma}$ defines the classical field amplitude.

We define the photon-number-weighted momentum sums
\begin{equation*}
M_n \equiv \sum_{\vec k,\sigma} \bigl\langle \hat n_{\vec k,\sigma}\bigr\rangle (\hbar k)^n , \qquad k \equiv |\vec k| ,
\end{equation*}
so that $M_1 \simeq \langle p_z \rangle$ for a directional pulse. Because the $M_n$ are built from the wave vector, the per-mode velocity must be taken in the wave-vector form derived above, $v_z(\vec k) = c(k_z/k)\bigl(1 - \alpha_0 l_\text{Pl} k + \beta_0^{(k)} l_\text{Pl}^2 k^2 - \cdots\bigr)$. We take the pulse velocity to be the energy-weighted mean of these per-mode velocities, the standard energy-transport definition and the quantity computed at zeroth order in Ref.~\cite{fedorov2017}. Because $\hbar\omega_k = \hbar c k$ exactly for photons, the factor $k$ in the energy weight cancels the $1/k$ in the per-mode velocity, and the only approximation entering is $k_z \simeq k$, valid for $w \gg \lambda$; no factorization of expectation values is assumed anywhere (see Appendix~\ref{app:influence} for the derivation). The result is
\begin{equation}
\langle v_{g} \rangle = v_{g,0}  \left( 1 - \alpha_0 \frac{ l_\text{Pl} }{\hbar} \frac{ M_2}{\langle p_z \rangle} + \beta_0^{(k)} \Bigl(\frac{ l_\text{Pl}}{\hbar}\Bigr)^{\!2} \frac{ M_3}{\langle p_z \rangle} - \cdots \right),
\label{eq:GUP_vg_macro}
\end{equation}
where $p \approx p_z$ is assumed for a highly directional pulse traveling along the $z$-axis, and $v_{g,0} = \langle p_z \rangle c^2/\langle E \rangle \approx c(1 - \frac{\lambda^2}{8\pi^2 w^2})$ is the unperturbed classical group velocity incorporating the geometric diffractive delay of a focused Gaussian beam with central wavelength $\lambda = 2\pi c / \omega_0$ and transverse beam waist $w \gg \lambda$~\cite{fedorov2017}.

The $M_n$ are extensive sums of the $n$-th power of the momenta of \textit{individual} photons (rather than moments of the macroscopic total momentum), so the extensive macroscopic parameters factor out of the ratios $M_n/M_1$.

By mapping the discrete quantum mode amplitudes to a classical Gaussian spatiotemporal field distribution with peak electric field strength ${\cal E}_0$ and pulse duration $\Delta t$, the momentum expectations under the short-pulse limit ($\omega_0 \Delta t \gg 1$) yield analytically (see Appendix~\ref{app:coherent})
\begin{eqnarray}
\langle p_z \rangle &\approx& \frac{\sqrt{\pi} \Delta t w^2 {\cal E}_0^2}{8}, \label{eq:pz} \\
M_2 &\approx& \frac{\sqrt{\pi} \Delta t w^2 {\cal E}_0^2 \hbar \omega_0}{8c}, \label{eq:p2} \\
M_3 &\approx& \frac{\sqrt{\pi} \Delta t w^2 {\cal E}_0^2 \hbar^2 \omega_0^2}{8c^2}. \label{eq:p3}
\end{eqnarray}
Equation~(\ref{eq:pz}) suppresses a multiplicative factor $\bigl(1-3\lambda^2/8\pi^2 w^2\bigr)$ derived in Appendix~\ref{app:coherent}; it cancels against the corresponding factor in $\langle E \rangle$ within $v_{g,0}$ and multiplies the already Planck-suppressed correction terms, where it is negligible.

We note that $M_3$ is the natural cubic moment for the quadratic GUP, and that no off-diagonal mode correlations enter (which is itself a consequence of the per-mode prescription). Substituting these moments into Eq.~(\ref{eq:GUP_vg_macro}), we see that the extensive parameter ${\cal E}_0^2 w^2 \Delta t$ (proportional to total pulse energy) scales homogeneously in both the numerator and denominator, thus canceling out. The GUP modification terms reduce to
\begin{eqnarray}
\alpha_0 \frac{ l_\text{Pl} }{\hbar} \frac{ M_2}{\langle p_z \rangle} &\approx& \alpha_0 l_\text{Pl} \frac{\omega_0}{c} = 2\pi \alpha_0 \frac{l_\text{Pl}}{\lambda}, \label{eq:cancel1} \\
\beta_0^{(k)} \Bigl(\frac{ l_\text{Pl}}{\hbar}\Bigr)^{\!2} \frac{ M_3}{\langle p_z \rangle} &\approx& \beta_0^{(k)} \frac{l_\text{Pl}^2 \omega_0^2}{c^2} = 4\pi^2 \beta_0^{(k)} \frac{l_\text{Pl}^2}{\lambda^2}. \label{eq:cancel2}
\end{eqnarray}

The effective group velocity of the macroscopic laser pulse is then
\begin{equation}
\langle v_{g} \rangle \approx v_{g,0} \left( 1 - 2\pi \alpha_0 \frac{l_\text{Pl}}{\lambda} + 4\pi^2 \beta_0^{(k)} \frac{l_\text{Pl}^2}{\lambda^2} \right).
\label{eq:GUP_vg_final}
\end{equation}
This confirms by explicit calculation what the QFT-natural prescription predicts: the macroscopic group velocity of a coherent pulse coincides with the classical pulse velocity computed from the GUP-modified single-photon velocity--momentum relation. We state the logical status of this result plainly: a per-mode deformation is assumed, and a per-photon answer is verified, so the absence of the soccer-ball problem holds by construction. The non-trivial content is that the ratio $M_2/M_1$ selects the central frequency $\omega_0$ rather than a bandwidth-weighted combination, and that no off-diagonal mode correlations enter. We note that at this order the result depends only on the spectral occupation $\langle \hat n_{\vec k,\sigma}\rangle$, so coherent, thermal, and Fock states with identical spectral intensity yield the same macroscopic group velocity; non-classical photon statistics do not enter at leading order. This differs from imposing the deformation on the canonical pair of a single harmonic oscillator, where the induced quartic term makes the spectrum quadratic in the excitation number and the Gazeau--Klauder coherent states of the oscillator become sub-Poissonian and exhibit fractional revivals~\cite{ghosh2012,iqbal20132013}. In the per-mode prescription the deformation enters only through the single-photon dispersion $E(k)$, so each mode keeps an equally spaced spectrum and a Glauber coherent state remains coherent and Poissonian under free propagation.

\begin{figure*}[t!]
\centering
\includegraphics[width=\linewidth]{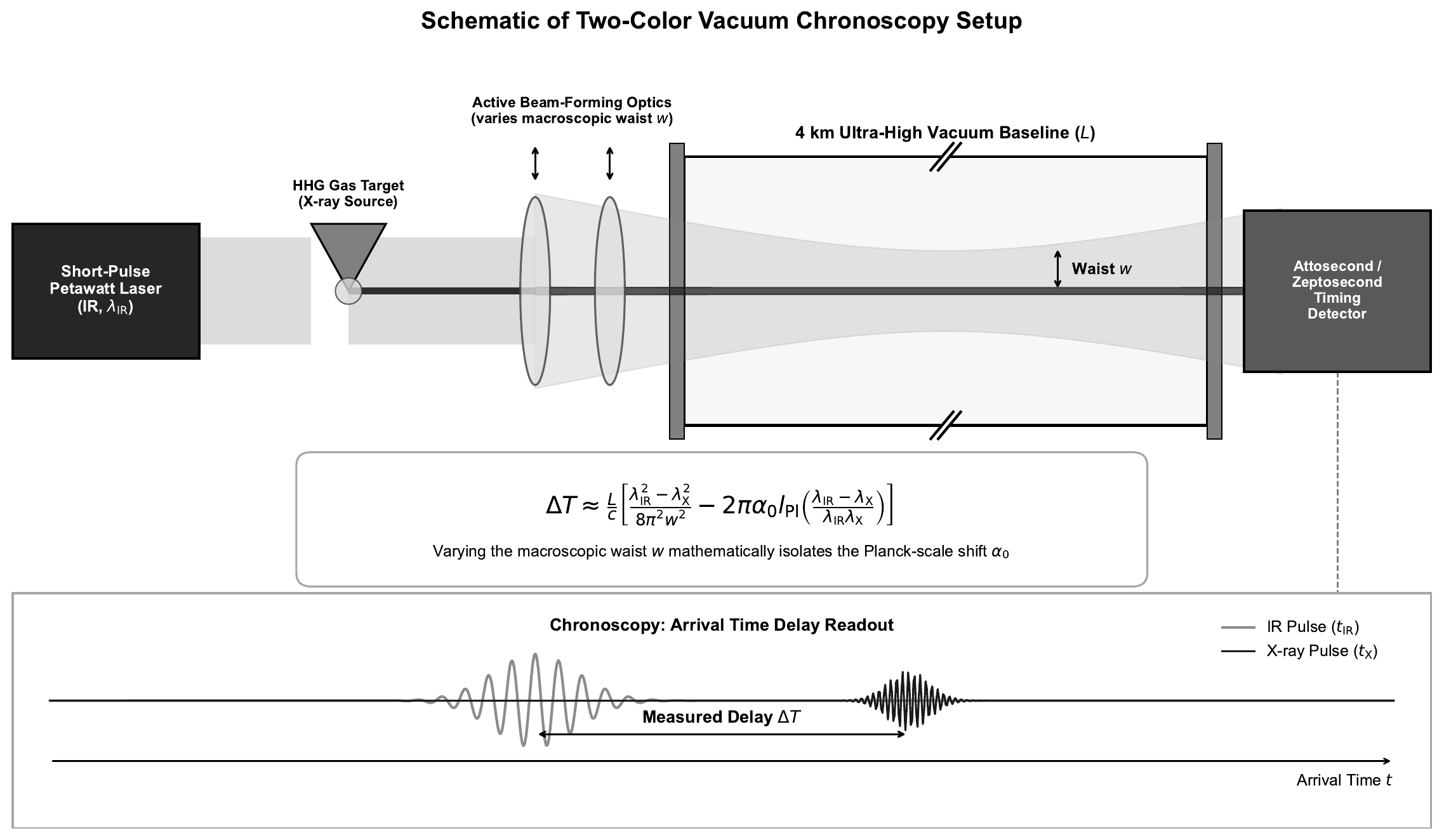}
\caption{\textbf{Schematic of the two-color vacuum chronoscopy experiment and isolation of Planck-scale kinematics.} The upper part shows the conceptual layout of the proposed terrestrial setup. A short-pulse petawatt laser generates an intense fundamental infrared (IR) wavepacket ($\lambda_\text{IR}$). Passing this beam through a high-harmonic generation (HHG) gas target creates a collinear, sub-cycle synchronized short-wavelength X-ray pulse ($\lambda_\text{X}$). The IR beam is drawn as the light gray envelope and the X-ray beam as the dark core on the axis. Active beam-forming optics systematically tune the macroscopic transverse beam waist $w$ before the co-propagating pulses travel through a $4\text{ km}$ ultra-high vacuum baseline ($L$). The lower panel shows the attosecond/zeptosecond resolution chronoscopy measurement, in which ultra-fast timing detectors at the baseline terminus measure the temporal delay $\Delta T$ between the IR (gray) and X-ray (black) wavepackets. Because the GUP correction is applied at the level of single-mode commutators, the macroscopic group velocity inherits the single-photon wavelength scale, and the chronoscopy addresses this kinematic shift without the macroscopic ``soccer-ball problem'' and without astrophysical source modeling.}
\label{fig:chronoscopy}
\end{figure*}

\section{Two-color vacuum chronoscopy}
\label{sec:chrono}

Because the macroscopic group velocity inherits the single-photon wavelength scale, increasing the total laser intensity does not magnify the physical velocity shift. We propose a ``two-color vacuum chronoscopy'' experiment that isolates this microscopic signature by exploiting the distinct spatial scaling behaviors of classical diffraction and quantum gravity.

The expected time-of-flight for a laser pulse propagating over a macroscopic ultra-high vacuum baseline $L$ is $t = L/\langle v_g \rangle$. Using Eq.~(\ref{eq:GUP_vg_final}), the time-of-flight to leading order is
\begin{equation}
t(\lambda, w) \approx \frac{L}{c} \left( 1 + \frac{\lambda^2}{8\pi^2 w^2} + 2\pi\alpha_0 \frac{l_\text{Pl}}{\lambda} - 4\pi^2\beta_0^{(k)} \frac{l_\text{Pl}^2}{\lambda^2} \right) .
\label{eq:ToF_Analytical}
\end{equation}

To detect the Planck-scale deformation, we evaluate the relative arrival time difference ($\Delta T$) between a fundamental optical/infrared laser pulse ($\lambda_\text{IR}$) and its synchronized High-Harmonic Generation (HHG) X-ray pulse ($\lambda_\text{X}$). Since the X-ray pulse is generated by the optical field inside a thin noble gas target, the two pulses share a common origin, separated only by the birth-time offset $\tau_0$ of the harmonic emission within the driving half-cycle. This offset is not small on the scale of interest, and it is waist-independent, so it survives the extrapolation below. It is, however, independent of the propagation length, whereas every term in Eq.~(\ref{eq:ToF_Analytical}) carries a factor $L/c$; we return to its removal below.

After co-propagating over the baseline $L$, assuming a linearly modified GUP model ($\alpha_0 \neq 0$) and $\lambda_\text{IR} \gg \lambda_\text{X}$, the arrival time delay $\Delta T = t(\lambda_\text{IR}, w) - t(\lambda_\text{X}, w)$ evaluates to
\begin{equation}
\Delta T \approx \frac{L}{c} \Biggl[ \frac{\lambda_\text{IR}^2 - \lambda_\text{X}^2}{8\pi^2 w^2}  - 2\pi\alpha_0 l_\text{Pl} \left( \frac{\lambda_\text{IR}-\lambda_\text{X}}{\lambda_\text{IR}\lambda_\text{X}} \right) \Biggr].
\label{eq:DeltaT_approx}
\end{equation}

\begin{figure}[t]
\centering
\includegraphics[width=\linewidth]{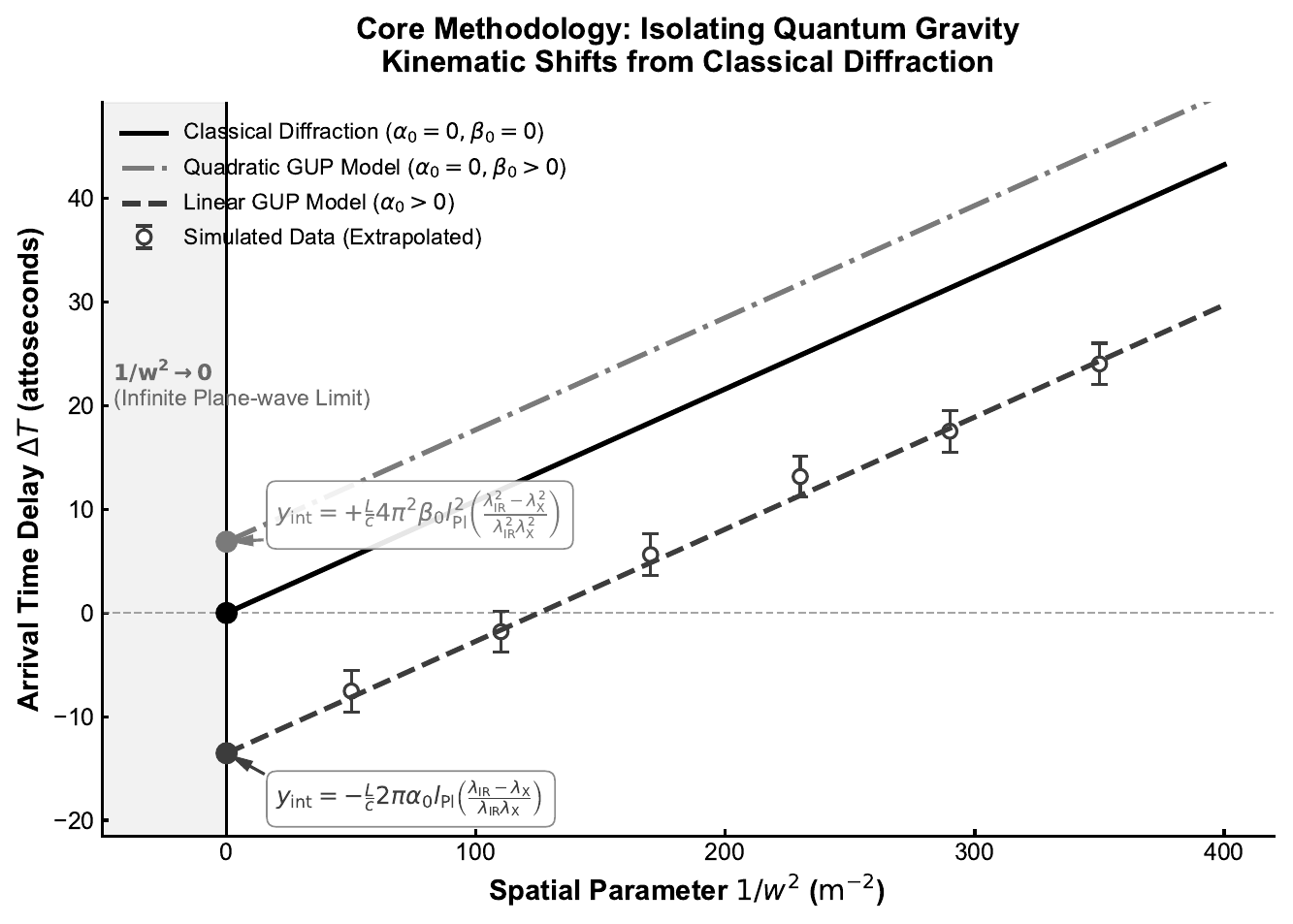}
\caption{\textbf{Isolation of Planck-scale kinematic shifts via active-waist vacuum chronoscopy.} Expected relative arrival time delay $\Delta T$ between a fundamental infrared laser pulse ($\lambda_\text{IR}$) and its synchronized high-harmonic X-ray pulse ($\lambda_\text{X}$) co-propagating over a macroscopic vacuum baseline, plotted as a function of the spatial geometric parameter $1/w^2$. The solid black line denotes the classical, subluminal geometric delay governed solely by macroscopic diffraction ($\alpha_0=0, \beta_0=0$). The dashed dark gray and dash-dotted light gray lines illustrate the theoretical macroscopic temporal trajectories introduced by the linear ($\alpha_0 > 0$) and quadratic ($\beta_0 > 0$) generalized uncertainty principle (GUP) modifications, respectively (GUP parameters are mathematically scaled here for visual clarity alongside the classical attosecond scale). The simulated data points (open circles, incorporating representative attosecond metrological error bars) illustrate the proposed experimental methodology. By systematically measuring temporal delays across actively varied finite macroscopic beam waists ($w$) and performing a linear regression, one can directly extrapolate to the infinite plane-wave limit ($1/w^2 \rightarrow 0$, shaded region). This limit removes the waist-dependent classical diffraction term; waist-independent systematics, principally residual-gas dispersion and the HHG birth-time offset (see text), remain in the intercept and must be removed before the residual intercept quantifies the single-photon GUP kinematic shift.}
\label{fig:methodology}
\end{figure}

Experimentally, as illustrated in Fig.~\ref{fig:chronoscopy}, the classical diffractive delay (the first term) can be disentangled from the GUP term. By systematically varying the macroscopic beam waist $w$ via active beam-forming optics and plotting the measured temporal delay $\Delta T$ against the spatial geometric parameter $1/w^2$, one observes a linear relationship, as shown in Fig.~\ref{fig:methodology}. The $y$-intercept of this plot (corresponding to the mathematical limit of an infinite plane wave, $1/w^2 \rightarrow 0$) removes the waist-dependent classical diffraction term. Any waist-independent systematic, however, survives at the intercept; two such terms, residual-gas dispersion and the harmonic birth-time offset, are quantified in Sec.~\ref{sec:feasibility} and must be removed independently. After that subtraction, any statistically significant, waist-independent negative delay at this intercept provides a direct measurement of the linear GUP parameter $\alpha_0$.

Similarly, for parity-preserving quantum gravity models where the linear parameter vanishes ($\alpha_0 = 0$), the macroscopic group velocity modification is governed by the quadratic parameter $\beta_0^{(k)}$, which in this case coincides with $\beta_0$. The temporal delay becomes
\begin{equation}
\Delta T \approx \frac{L}{c} \Biggl[ \frac{\lambda_\text{IR}^2 - \lambda_\text{X}^2}{8\pi^2 w^2} + 4\pi^2 \beta_0^{(k)} l_\text{Pl}^2 \left( \frac{\lambda_\text{IR}^2 - \lambda_\text{X}^2}{\lambda_\text{IR}^2 \lambda_\text{X}^2}  \right) \Biggr] .
\label{eq:DeltaT_beta}
\end{equation}
For $\beta_0 > 0$, the quadratic GUP correction increases the group velocity at high energies, giving a kinematic superluminal advance that opposes the subluminal classical geometric delay.

The sign of the predicted shift differs between the two GUP models, providing an experimental discriminator. The classical diffractive delay and the linear $\alpha_0 > 0$ correction both contribute subluminally, in the same direction. By contrast, the quadratic $\beta_0 > 0$ correction acts \emph{superluminally} on the high-energy pulse, producing a kinematic advance that opposes the diffractive delay. A measurement of the sign of the waist-independent intercept therefore distinguishes parity-violating ($\alpha_0$) from parity-preserving ($\beta_0$) GUP models, independently of the magnitude of the effect. We caution that residual-gas dispersion delays the infrared pulse relative to the X-ray pulse and therefore enters the intercept with the same sign as the $\beta_0>0$ advance while partially canceling the $\alpha_0>0$ delay: the sign discrimination is meaningful only after the gas term has been characterized and subtracted, as discussed below.

\section{Observational baseline and feasibility}
\label{sec:feasibility}

A terrestrial null result at zeptosecond timing precision would yield $\alpha_0 \lesssim 7.4\times10^{8}$ (Appendix~\ref{app:estimate}) and $\beta_0^{(k)} \lesssim 7\times 10^{33}$: the quadratic GUP advance over the baseline is $(L/c)\,4\pi^2\beta_0^{(k)}\,l_{\rm Pl}^2/\lambda_{\rm X}^2 \approx 1.4\times10^{-55}\,\beta_0^{(k)}$~s, so the method has no meaningful quadratic reach at foreseeable timing precision, and the quadratic term is discussed above only as a sign discriminator in principle. These bounds are numerically far weaker than published time-of-flight constraints from gamma-ray bursts. Those constraints are stated as a velocity $v = c[1 - s\,\tfrac{n+1}{2}(E/E_{{\rm QG},n})^n]$ in the photon energy $E$, with $s = +1$ ($-1$) for the subluminal (superluminal) branch. Our $\alpha_0 > 0$ term is subluminal and translates with unit convention factor, $\alpha_0 = E_{\rm Pl}/E_{{\rm QG},1}$; our $\beta_0 > 0$ term is superluminal and requires the superluminal quadratic branch, $\beta_0 = \tfrac{3}{2}(E_{\rm Pl}/E_{{\rm QG},2})^2$. Because these limits are expressed in $E = cp$, they constrain $\beta_0$ rather than $\beta_0^{(k)}$. On this basis GRB~090510~\cite{vasileiou2013} gives $\alpha_0 \lesssim 0.13$ and $\beta_0 \lesssim 2.5\times10^{16}$. The LHAASO observations of GRB~221009A~\cite{lhaaso2024} quote $E_{{\rm QG},1} > 10\,E_{\rm Pl}$ and $E_{{\rm QG},2} > 6\times10^{-8}E_{\rm Pl}$ without separating the branches, giving $\alpha_0 \lesssim 0.10$ and $\beta_0 \lesssim 4.2\times10^{14}$; the branch-resolved analysis of the same dataset in Ref.~\cite{yang2024} gives $\alpha_0 \lesssim 0.083$ for the subluminal linear case and $\beta_0 \lesssim 4.3\times10^{14}$ for the superluminal quadratic case, which are the branches relevant here. Independent analyses of that dataset span roughly a factor of two. We note that the strongest of these limits are obtained by fitting an assumed intrinsic light curve and spectral evolution at the source, which is the modeling a terrestrial measurement is designed to avoid; under the maximally conservative treatment of source-intrinsic spectral evolution in Ref.~\cite{vasileiou2013} the linear limit degrades only to $E_{{\rm QG},1} > 2E_{\rm Pl}$, i.e.\ $\alpha_0 \lesssim 0.5$, so the astrophysical systematics cost the cosmological bound a factor of a few rather than orders of magnitude, and we make no claim that the proposed setup is currently competitive at the Planck scale. Reaching $\alpha_0 \sim \mathcal{O}(1)$ at $L = 4$~km and $\lambda_X = 1$~nm would require timing precision $\tau_{\mathrm{res}} \sim 10^{-30}$~s, approximately nine orders of magnitude beyond present capability. Achieving zeptosecond timing precision requires sufficient photon flux at the detector after the 4~km baseline; for soft X-ray HHG sources at $\lambda_X \approx 1$~nm, this sets a practical floor on integration time and informs the choice of repetition rate.

The first of two waist-independent systematics is residual-gas dispersion (Appendix~\ref{app:gas}). At $10^{-9}$~torr the group refractivity of hydrogen-dominated residual gas delays the infrared pulse by $\approx 2.3\times10^{-21}$~s over $L=4$~km ($\approx 4.9\times10^{-21}$~s for air-like composition), while the X-ray group delay is four orders of magnitude smaller. This term exceeds the assumed timing floor, is independent of $w$, and enters $\Delta T$ with the same sign as a $\beta_0>0$ advance; left uncharacterized it would limit the bound to $\alpha_0 \lesssim 1.7\times10^{9}$ and defeat the sign discrimination. It cannot be removed by backfilling a test gas and regressing on the added pressure, because the pre-existing residual sits at fixed partial pressure across such a scan and therefore contributes a constant that is indistinguishable from the intercept. What does work is to vary the base pressure itself at fixed composition, by throttling the pumping speed, and to include the measured total pressure as a second regression variable alongside $1/w^2$. Equivalently, residual-gas-analyzer characterization of the column to better than $\sim 40\%$ suppresses the subtraction residual below $10^{-21}$~s, and operating at $P \lesssim 4\times10^{-10}$~torr of H$_2$ renders the term subdominant outright. The bound quoted above assumes one of these; without it the experiment is systematics-limited rather than timing-limited. The second waist-independent term is the HHG birth-time offset $\tau_0$ (Appendix~\ref{app:tau0}), which is of order $10^{-16}$--$10^{-15}$~s and therefore far larger, but which is removed without needing to be known: $\tau_0$ is independent of the propagation length while the diffractive, residual-gas and GUP contributions to $\Delta T$ are all proportional to $L$, so repeating the waist regression at two or more baselines and extrapolating the resulting intercepts linearly in $L$ separates them.

Existing kilometer-scale ultra-high-vacuum facilities developed for gravitational-wave interferometry provide a natural infrastructure baseline, though a dedicated installation would have very different optical requirements and the residual-gas pressure and composition requirements above as added specifications. Other terrestrial probes of GUP phenomenology, including optomechanical proposals~\cite{pikovski2012} and high-precision spectroscopy~\cite{das2011}, constrain different sectors of the deformation; the dispersion-sector bound established here is complementary. The value of the proposal is not in immediate sensitivity but in establishing a controlled, source-independent terrestrial methodology, with a clear roadmap for improvement: (i) longer baselines $L$, (ii) shorter probe wavelengths $\lambda_X$ accessible through higher-order HHG or free-electron-laser sources, and (iii) quantum-enhanced timing techniques. The combination $L/\lambda_X$ is the relevant figure of merit and improves linearly under each.

\section{Conclusion}
\label{sec:conclusion}

We have shown by explicit calculation that, under the QFT-natural prescription in which the GUP modifies single-mode commutators, the macroscopic group velocity of a multimode coherent pulse reduces to the classical pulse velocity evaluated with the GUP-modified single-photon velocity--momentum relation at the central wavelength. The macroscopic energy enters numerator and denominator homogeneously and drops out. The soccer-ball problem, which arises only when the GUP is applied to the center-of-mass momentum of a composite system, is therefore absent for coherent light, and macroscopic petawatt pulses probe the same Planck-scale parameter space as single high-energy astrophysical photons. Since the commutator and momentum-space routes are related by an exact change of variable, the single-photon input is equivalent to a modified dispersion relation~\cite{amelino1998}, and the contribution of the calculation is the controlled passage from the single-photon deformation to the macroscopic coherent-state observable, not the single-photon relation itself.

Our proposed two-color vacuum chronoscopy uses active-waist optics to separate the classical diffractive delay from the intrinsic quantum gravitational signatures, combined with pressure modulation to control the residual-gas dispersion that would otherwise dominate the waist-independent intercept. This decouples the search for Planck-scale geometry from the complex emission models of unpredictable cosmic transients, establishing a clear experimental pathway for tests of spacetime geometry in controlled large-scale terrestrial laboratories.

At the order computed here, the multimode coherent-state expectation is sensitive only to the single-photon number distribution $\langle \hat n_{\vec k,\sigma}\rangle$, and the result coincides with what one would obtain from classical electromagnetism with GUP-modified dispersion. Genuinely non-classical signatures would require observables that couple to higher photon-number moments, for example two-photon arrival-time correlations in squeezed or sub-Poissonian light, where the variance of $\hat n_{\vec k,\sigma}$ enters at leading order. Such observables would constitute a distinctively quantum probe of GUP phenomenology and we identify them as a natural next step.

\section*{Data availability}
The data that support the findings of this article are not publicly available because they consist solely of numerical values generated directly from the analytical expressions and parameters given in the text. 

\begin{acknowledgments}
	S.D. acknowledges support from the Natural Sciences and Engineering Research Council of Canada (NSERC). The authors used Claude Opus 5.5 (Anthropic) to check selected algebraic and numerical steps and to improve the language of the manuscript. All derivations and numerical results were verified independently by the authors, who take full responsibility for the content of this article.
\end{acknowledgments}

\appendix

\section{Generalized uncertainty principle and the single-particle group velocity}
\label{app:gup}

Heisenberg's uncertainty principle can be stated
(in one dimension) as
\begin{eqnarray}
[x_0,p_0]=i\hbar \;\;\; \Leftrightarrow \;\;\;
\Delta x_0 \Delta p_0 \geq \frac{\hbar}{2} ,
\end{eqnarray}
where $p_0$ is the canonical momentum.

String theory, loop quantum gravity, doubly special relativity, black hole physics etc. all
imply a minimum measurement length $l_\text{min}$. To incorporate this minimum length, a generalized
uncertainty principle for the physical location and momentum is proposed
in the form~\cite{adv}
\begin{eqnarray}
[x,p] \!=\! i\hbar \Bigl( 1 - \alpha_0 \frac{ l_\text{Pl} }{\hbar} \, p
+ \beta_0 \Bigl(\frac{ l_\text{Pl}}{\hbar}\Bigr)^{\!2}  p^2 - \cdots
\Bigr) .
\label{GUncertainty}
\end{eqnarray}
Here $x=x_0$ is the physical position and $p$ is the physical momentum. We take Eq.~(\ref{GUncertainty}) as a generic phenomenological parameterization of linear-plus-quadratic minimal-length deformations~\cite{kempf1995,adv}, with $\alpha_0$ and $\beta_0$ treated as independent parameters. Their normalization is fixed by Eq.~(\ref{GUncertainty}) and therefore refers to an expansion in the \emph{physical} momentum $p$; this distinction matters at quadratic order and is made explicit in the paragraph following Eq.~(\ref{Gvelo1}). In the specific three-dimensional algebra of Ref.~\cite{adv}, $[x_i,p_j] = i\hbar\bigl[\delta_{ij} - \varepsilon( p\,\delta_{ij} + p_ip_j/p ) + \varepsilon^2( p^2\delta_{ij} + 3p_ip_j )\bigr]$ with $\varepsilon = \alpha_0^{\rm ADV} l_{\rm Pl}/\hbar$, the one-dimensional reduction along the propagation axis ($i=j=z$, $\vec p = p\hat z$) reads $[x,p] = i\hbar(1 - 2\varepsilon p + 4\varepsilon^2 p^2)$, so that $\alpha_0 = 2\alpha_0^{\rm ADV}$ and $\beta_0 = 4(\alpha_0^{\rm ADV})^2 = \alpha_0^2$ are tied together rather than independent. All bounds quoted in this work apply to the generic parameterization and specialize to any particular model by matching conventions.

If we label the four coordinates as $x^\mu = \{ct,x^1,x^2, x^3\}$. For a
trajectory in a flat spacetime (which can be described using special relativity),
the proper
length element $ds$ of this trajectory obeys the following relation
\begin{eqnarray}
ds^2 = -\eta_{\mu\nu} dx^\mu dx^\nu  = c^2dt^2 - (dx^i)^2.
\end{eqnarray}
The tangent vector of this trajectory may be written
\begin{eqnarray}
s^\mu = \frac{d x^\mu}{ds} = \frac{d x^\mu}{c d\tau} = \frac{1}{c} v^\mu,
\end{eqnarray}
where $d\tau$ is the proper time element and $c\, d\tau = ds$. Then, we have
\begin{eqnarray}
v^\mu v_\mu = c^2 s^\mu s_\mu = - c^2.
\end{eqnarray}
The momentum of a particle (system) traveling along this trajectory
may be written $p^\mu = v^\mu \mu c$, and having
\begin{eqnarray}
p^\mu p_\mu = - \mu^2 c^4.
\label{standardDis}
\end{eqnarray}
Then the energy and momentum has the following relations
\begin{eqnarray}
&&p^0 = \gamma \mu c^2 ~, \;\;\;\;\;\;\;\;\;\;\;\;\;\;\;\;\;\;
\vec p =  \gamma \mu \vec v ~, \label{em1} \\
&&p^\mu = \left(E,\vec p c \right)~, \;\;\;\;\;\;\;\;\;\;
E = \gamma \mu c^2~, \label{em2}
\end{eqnarray}
where $\gamma=1/\sqrt{1-v^2/c^2}$ and $\vec v=\frac{d\vec{x}}{dt}$.
In addition, Eqs.~(\ref{em1}) and (\ref{em2}) imply \cite{okun1989}
\begin{eqnarray}
\vec{v}= \frac{d \vec{x}}{dt} =\frac{\vec{p} c^2}{E} .
\label{velocity}
\end{eqnarray}
The standard dispersion relation Eq.~(\ref{standardDis}) holds
\begin{eqnarray}
p_\mu p^\mu = - E^2 + p^2 c^2 = - \mu^2 c^4~,
\label{disp1}
\end{eqnarray}
which together with Eq.\eqref{em1} implies that energy and momentum
simply add up for non-interacting particles. However, we will show that
this is not true for interacting particles in Appendix~\ref{app:coherent}. Such
kind of effect has been discussed before as the non-standard
energy-momentum dispersion relation and may be resolved by the
generalized uncertainty principle, see Ref.~[\onlinecite{bosso2019}] for
more details.

We may also calculate the group velocity of a
particle or a small system by the Poisson bracket and Hamiltonian formalism.

The velocity may be written as
\begin{eqnarray}
v_{g,0}&=& \frac{d x_0}{d t}  =  \frac{1}{i \hbar} [  x_0, H(p_0) ]
+ \frac{\partial  x_0}{\partial t }  \nonumber \\
&=&  \frac{1}{i \hbar} [  x_0, H(p_0) ]  \nonumber \\
&=& \frac{1}{i \hbar} \Bigl( \frac{ \partial}{\partial p_0 } H(p_0) \Bigr)  [  x_0, p_0 ]
\label{Ha21}
\\
&=& \frac{ \partial}{\partial p_0 } E(p_0)
=  \frac{ \partial}{\partial p_0 } \sqrt{p_0^2 c^2 + \mu^2 c^4}  =\frac{ p_0 c^2}{E} ,
\label{Ha2}
\end{eqnarray}
where $x_0$ does not depend explicitly on time,
the Hamiltonian is equated to
the
relativistic energy
in going from the third to
the fourth line. Therefore, when using the canonical Hamiltonian and
momentum, Eq.~(\ref{Ha2}) exactly agrees with the prediction of the
special relativity Eq.~(\ref{velocity}).

However, if we insert the physical momentum predicted by the GUP
Eqs.~(\ref{GUncertainty}) into (\ref{Ha21}), we will obtain a
different group velocity expression with higher-order corrections from quantum
gravity:
\begin{eqnarray}
v_g &=& \frac{d  x}{d t}  =  \frac{1}{i \hbar} [  x, H(p) ]
+ \frac{\partial  x}{\partial t }  \nonumber \\
&=& \frac{1}{i \hbar} \Bigl( \frac{ \partial}{\partial p } H(p) \Bigr) [ x, p ] \nonumber \\
&=& \frac{1}{i \hbar} \frac{ \partial}{\partial p } \sqrt{p^2 c^2 + \mu^2 c^4}  [ x, p ]
\nonumber \\
&=& \frac{ pc^2}{E} \left( 1 - \alpha_0 \frac{ l_\text{Pl} }{\hbar} \, p
+ \beta_0 \Bigl(\frac{ l_\text{Pl}}{\hbar}\Bigr)^{\!2}  p^2 - \cdots
\right) ,
\label{Gvelo1}
\end{eqnarray}
where we have used Eq.~(\ref{GUncertainty}) in going from the third
to the fourth line. In Appendices~\ref{app:coherent} and \ref{app:influence} we use
Eq.~(\ref{Gvelo1}) to calculate the influence of quantum gravity on the
group velocity of the multiple modes light pulses.

Because photons admit no position operator in the Newton--Wigner sense, the commutator route above cannot be applied to light as it stands. The deformation can, however, be stated entirely in momentum space, without any position operator entering. Let $p_0 = \hbar k$ denote the translation generator, which photons possess irrespective of their localization properties, and write the physical momentum as a map $p(p_0)$. Since $[x,p_0] = i\hbar$, the deformed algebra is $[x,p] = i\hbar\,dp/dp_0$, and the choice
\begin{eqnarray*}
p = p_0 \Bigl[ 1 - \tfrac{1}{2}\alpha_0 \tfrac{l_\text{Pl}}{\hbar} p_0 + \bigl(\tfrac{1}{3}\beta_0 + \tfrac{1}{6}\alpha_0^2\bigr) \Bigl(\tfrac{l_\text{Pl}}{\hbar}\Bigr)^{\!2} p_0^2 - \cdots \Bigr]
\end{eqnarray*}
reproduces Eq.~(\ref{GUncertainty}) exactly, order by order in $l_\text{Pl}$ and for arbitrary $\alpha_0$ and $\beta_0$. To verify this, invert the series to $p_0 = p + \tfrac{1}{2}\alpha_0(l_\text{Pl}/\hbar)p^2 + \tfrac{1}{3}(\alpha_0^2 - \beta_0)(l_\text{Pl}/\hbar)^2 p^3$ and substitute into $dp/dp_0 = 1 - \alpha_0 (l_\text{Pl}/\hbar) p_0 + (\beta_0 + \tfrac{1}{2}\alpha_0^2)(l_\text{Pl}/\hbar)^2 p_0^2$, which returns $1 - \alpha_0 (l_\text{Pl}/\hbar) p + \beta_0 (l_\text{Pl}/\hbar)^2 p^2$. The energy $E(k) = c\,p(k)$ is then a deformed function of the wave vector, and the group velocity $d\omega/dk = c\,dp/dp_0$ is well defined for a wave:
\begin{eqnarray*}
v_g &=& c \Bigl( 1 - \alpha_0 l_\text{Pl} k + \beta_0^{(k)} l_\text{Pl}^2 k^2 - \cdots \Bigr), \\
\beta_0^{(k)} &\equiv& \beta_0 + \tfrac{1}{2}\alpha_0^2 .
\end{eqnarray*}
This is Eq.~(\ref{Gvelo1}) rewritten in the wave-vector variable, using $pc^2/E = c$ for a massless mode. The two routes are therefore related by an exact change of variable and no downstream result depends on which is used, provided the same variable is used throughout; the wave-mechanical route is the logically primary one for light. The change of variable is not innocuous at second order: the linear coefficient is unchanged, but $\beta_0$ is the quadratic coefficient of the velocity as a function of the physical momentum, equivalently of the photon energy $E = cp$, whereas $\beta_0^{(k)}$ is its coefficient as a function of the wave vector, equivalently of the wavelength $\lambda = 2\pi/k$. This matters below because the mode sums are built from $\hbar k$ while the astrophysical time-of-flight limits are quoted in $E$. The two coincide whenever $\alpha_0 = 0$. In the algebra of Ref.~\cite{adv}, where $\alpha_0 = 2\alpha_0^{\rm ADV}$ and $\beta_0 = \alpha_0^2$, this reproduces the familiar statement that the quadratic coefficient of the velocity reads $6(\alpha_0^{\rm ADV} l_\text{Pl}/\hbar)^2$ in the variable $\hbar k$ and $4(\alpha_0^{\rm ADV} l_\text{Pl}/\hbar)^2$ in the physical momentum. This equivalence also makes precise the relation to the modified-dispersion-relation literature~\cite{amelino1998}: the single-photon input used below is a modified dispersion relation, and the calculation of Appendices~\ref{app:coherent} and \ref{app:influence} should be read as a controlled passage from that per-mode deformation to the macroscopic coherent-state observable, a consistency check of the per-mode prescription, rather than as a new single-photon prediction.

\section{Energy and momentum moments of multimode coherent light pulses}
\label{app:coherent}

We now consider the group velocity of multimode coherent light pulses.
We calculate the group velocity of multimode coherent light pulses following
Ref.~[\onlinecite{fedorov2017}].

\subsection{Quantum description}

In quantum mechanics, a light pulse may be described as a multimode coherent state:
\begin{eqnarray}
\vert{\Psi}\rangle
= \prod_{\vec{k},\sigma} e^{-\frac{|\alpha_{\vec{k},\sigma}|^2}{2}} \sum_{n_{\vec{k},\sigma}} \frac{(\alpha_{\vec{k},\sigma} a_{\vec{k},\sigma}^\dagger)^{n_{\vec{k},\sigma}}}{n_{\vec{k},\sigma} ! } \vert{0}\rangle ,
\label{state}
\end{eqnarray}
where $\vert{0}\rangle$ is the ground vacuum state, $a_{\vec{k},\sigma}^\dagger$ is the raising (or creation) operator for modes with momentum $\vec{k}$ and polarization $\sigma$, and $|\alpha_{\vec{k},\sigma}|^2$ describes photon numbers in the specific mode with momentum $\vec{k}$ and polarization $\sigma$.

Therefore, the total number of photons in all modes may be written as
\begin{eqnarray}
N = \sum_{\vec{k},\sigma} |\alpha_{\vec{k},\sigma}|^2 = \sum_{\sigma}  \frac{V}{(2 \pi)^3} \int |\alpha_{\vec{k},\sigma}|^2 d \vec{k} .
\end{eqnarray}
Recall that for each single photon, we have $p^\mu=(\frac{\hbar \omega_k}{c}, \hbar \vec{k})$ with $p^\mu p_\mu=0$ and hence $\omega_k= c |\vec{k}|$. The energy of the multimode quantum state described by Eq.~(\ref{state}) may be calculated as
\begin{eqnarray}
\langle E \rangle &=& \sum_{\vec{k},\sigma} \hbar \omega_k
\langle\Psi\vert{a_{\vec{k},\sigma}^\dagger a_{\vec{k},\sigma}}
\vert{\Psi}\rangle = \sum_{\vec{k},\sigma} \hbar \omega_k  |\alpha_{\vec{k},\sigma}|^2 \nonumber \\
&=& \sum_{\sigma} \frac{V}{(2 \pi)^3} \int \hbar \omega_k |\alpha_{\vec{k},\sigma}|^2 d \vec{k} .
\label{Qenergy}
\end{eqnarray}
Similarly, the momentum of the light pulses equal
\begin{eqnarray}
\langle \vec{p} \rangle &=& \sum_{\vec{k},\sigma} \hbar \vec{k}\,
\langle\Psi\vert{a_{\vec{k},\sigma}^\dagger a_{\vec{k},\sigma}}
\vert{\Psi}\rangle = \sum_{\vec{k},\sigma} \hbar \vec{k}  |\alpha_{\vec{k},\sigma}|^2 \nonumber \\
&=& \sum_{\sigma} \frac{V}{(2 \pi)^3} \int \hbar \vec{k} \, |\alpha_{\vec{k},\sigma}|^2 d \vec{k} ,
\label{Qmomentum}
\end{eqnarray}
We further define the photon-number-weighted momentum sums
\begin{eqnarray*}
M_n \equiv \sum_{\vec{k},\sigma} \bigl\langle a_{\vec{k},\sigma}^\dagger a_{\vec{k},\sigma} \bigr\rangle (\hbar k)^n , \qquad k \equiv |\vec k| ,
\end{eqnarray*}
which are extensive sums of the $n$-th powers of the momenta of \emph{individual} photons, not moments of the total pulse momentum, so that $M_1 \simeq \langle p_z \rangle$ for a directional pulse. In particular
\begin{eqnarray}
M_2 &=& \sum_{\vec{k},\sigma} \hbar^2 (\vec{k})^2\,
\langle\Psi\vert{a_{\vec{k},\sigma}^\dagger a_{\vec{k},\sigma}}
\vert{\Psi}\rangle
= \sum_{\vec{k},\sigma} \hbar^2 \frac{\omega_k^2}{c^2}  |\alpha_{\vec{k},\sigma}|^2 \nonumber \\
&=& \sum_{\sigma} \frac{V}{(2 \pi)^3} \int  \frac{\hbar^2 \omega_k^2}{c^2} \, |\alpha_{\vec{k},\sigma}|^2 d \vec{k} ,
\label{QmomentumSquare}
\end{eqnarray}
where $M_2$ calculates the sum of squared momenta of individual photons, not the square of the overall momentum of the macroscopic pulse.

Similarly, the expectation value of the cubed momentum of the individual photons may be calculated as
\begin{eqnarray}
M_3 \!\!\!&=&\!\!\! \sum_{\vec{k},\sigma} \hbar^3 |\vec{k}|^3\,
\langle\Psi\vert{a_{\vec{k},\sigma}^\dagger a_{\vec{k},\sigma}}
\vert{\Psi}\rangle \nonumber \\
\!\!\!&=&\!\!\! \sum_{\vec{k},\sigma} \hbar^3 \frac{\omega_k^3}{c^3}  |\alpha_{\vec{k},\sigma}|^2 \nonumber \\
\!\!\!&=&\!\!\! \sum_{\sigma} \frac{V}{(2 \pi)^3} \int  \frac{\hbar^3 \omega_k^3}{c^3} \, |\alpha_{\vec{k},\sigma}|^2 d \vec{k} ,
\label{QmomentumCube1}
\end{eqnarray}
where $M_3$ evaluates the sum of the cubed momenta of individual photons.

Inserting Eqs.~(\ref{Qenergy}) and (\ref{Qmomentum}) into the energy-momentum relation (dispersion relation) Eq.~(\ref{disp1}) yields
\begin{eqnarray}
\mu^2 c^4 &=& \langle E \rangle^2 - c^2 \langle \vec{p} \rangle^2 \nonumber \\
&=& - c^2 \sum_{\vec{k}, \vec{k}'} \bigl( p^\mu p'_\mu \sum_{\sigma,\sigma'} |\alpha_{\vec{k},\sigma}|^2 |\alpha_{\vec{k}',\sigma'}|^2 \bigr) .
\label{NewE}
\end{eqnarray}

Similarly, the averaged electric field strength may be calculated
\begin{eqnarray}
&&\!\!\!\!\!\!\!\!\!\!\!\! \langle {\cal E} \rangle_\sigma(t,\vec{r}) = \langle\Psi\vert{ {\cal E}_\sigma}\vert{\Psi}\rangle \nonumber \\
&=& i \sum_{\vec{k}}  \sqrt{\frac{2\pi\hbar\omega_k}{V}} |\alpha_{\vec{k},\sigma}| \nonumber \\
&& \times \Bigl( e^{i(\vec{k}\cdot \vec{r} - \omega_k t + \varphi_{\vec{k},\sigma})} - e^{-i(\vec{k}\cdot \vec{r} - \omega_k t + \varphi_{\vec{k},\sigma})} \Bigr) \nonumber \\
&=&\!\!\! -\frac{1}{2^{3/2} \pi^{5/2}} \int \!\!\! \sqrt{V \hbar \omega_k} |\alpha_{\vec{k},\sigma}| \sin(\vec{k}\cdot \vec{r} \!-\! \omega_k t \!+\! \varphi_{\vec{k},\sigma}) d \vec{k} \nonumber \\
&=&\!\!\! \frac{1}{2^{3/2} \pi^{5/2}} \int \!\!\! \sqrt{V \hbar \omega_k} |\alpha_{\vec{k},\sigma}| \sin(\omega_k t \!-\! \vec{k}\cdot \vec{r} \!-\! \varphi_{\vec{k},\sigma}) d \vec{k},\nonumber \\
\label{QelctS}
\end{eqnarray}
where $\varphi_{\vec{k},\sigma}$ is the phase of the complex parameter $\alpha_{\vec{k},\sigma}$ and $ {\cal E}_\sigma$ is defined as
\begin{eqnarray}
{\cal E}_\sigma \!\!=\!\! i \sum_{\vec{k}} \!\! \sqrt{\frac{2\pi\hbar\omega_k}{V}} \Bigl(\!  a_{\vec{k},\sigma} e^{i(\vec{k}\cdot \vec{r}- \omega_k t)} \!-\! a_{\vec{k},\sigma}^\dagger e^{-i(\vec{k}\cdot \vec{r} - \omega_k t)} \!\Bigr). \nonumber \\
\label{app:Efield}
\end{eqnarray}

There is considerable
freedom in the choice of the parameters in the quantum description of the light pulses, to fix which, we now consider the classical description of the multimode coherent light pulses.

\subsection{Classical description}

In a classical description, it is reasonable to assume that both the frequency distribution and the temporal shape of the light pulses obey the Gaussian distribution.
Assuming that the light pulses are traveling towards the
right along the $z$-axis
and the plane orthogonal to this is the $xy$-plane, the electric field strength at $z=0$ may be written as
\begin{eqnarray}
{\cal E}_\sigma(t,\vec{r}_\perp)|_{z=0}  \!\!\!&=&\!\!\! {\cal E}_\sigma^{(0)} (\vec{r}_\perp) \sin(\omega_0 t) e^{-\frac{t^2}{2\Delta  t^2}}  \nonumber \\
&=&\!\!\! {\cal E}_\sigma^{(0)} (\vec{r}_\perp) \frac{i}{2} (e^{-i \omega_0 t}-e^{i \omega_0 t}) e^{-\frac{t^2}{2\Delta  t^2}} ,
\label{CEstren}
\end{eqnarray}
where $\omega_0$ is the mean frequency of the field and $\Delta  t$ is the light pulse duration. Before continuing our analysis, let us first review the Fourier transformation of $e^{-i \omega_0 t}e^{-\frac{t^2}{2\Delta  t^2}}$ in Eq.~(\ref{CEstren})
\begin{eqnarray}
&& \frac{\Delta t}{\sqrt{2\pi}} \int_{-\infty}^{+\infty} e^{i\omega t} e^{-\frac{(\omega+\omega_0)^2 \Delta  t^2}{2}} d\omega \nonumber \\
&=& \frac{1}{\sqrt{2\pi}} \int_{-\infty}^{+\infty} e^{-i\omega_0 t} e^{i(\omega + \omega_0) t} e^{-\frac{(\omega+\omega_0)^2 \Delta  t^2}{2}} d(\Delta t \, (\omega + \omega_0)) \nonumber \\
&=& e^{-i\omega_0 t} \frac{1}{\sqrt{2\pi}} \int_{-\infty}^{+\infty} e^{i(\omega + \omega_0)\Delta t \frac{t}{\Delta t}} e^{-\frac{(\omega+\omega_0)^2 \Delta  t^2}{2}} d(\Delta t \, (\omega + \omega_0))  \nonumber \\
&=& e^{-i\omega_0 t} \frac{1}{\sqrt{2\pi}} \int_{-\infty}^{+\infty} e^{i \, s \frac{t}{\Delta t}} e^{-\frac{s^2}{2}} ds \nonumber \\
&=&  e^{-i\omega_0 t} e^{-\frac{t^2}{2 \Delta t^2}} \, ,
\label{Fouri1}
\end{eqnarray}
where we have replaced $\Delta t \, (\omega + \omega_0)$ by $s$ in going from the third to the fourth line.
Similarly, we have
\begin{eqnarray}
\frac{\Delta t}{\sqrt{2\pi}} \int_{-\infty}^{+\infty} e^{i\omega t} e^{-\frac{(\omega-\omega_0)^2 \Delta  t^2}{2}} d\omega
= e^{i\omega_0 t} e^{-\frac{t^2}{2 \Delta t^2}}  \, .
\label{Fouri2}
\end{eqnarray}
Using the results of Eqs.~(\ref{Fouri1}) and (\ref{Fouri2}), the electric field strength in Eq.~(\ref{CEstren}) may be written as the Fourier transform with respect to $\vec{k}_\perp$ and the frequency distribution $~~~\omega$
as follows
\begin{widetext}
\begin{eqnarray}
{\cal E}_\sigma(t,\vec{r}_\perp)|_{z=0} \!\!\!&=&\!\!\!  \frac{i}{2} \frac{1}{(\sqrt{2\pi})^2} \int e^{i\vec{k}_\perp \cdot \vec{r}_\perp}  \tilde{{\cal E}}_\sigma^{(0)} (\vec{k}_\perp) d\vec{k}_\perp
\frac{\Delta  t}{\sqrt{2\pi}} \int_{-\infty}^{+\infty} e^{i \omega t} \left(e^{-\frac{(\omega+\omega_0)^2 \Delta  t^2}{2}} -e^{-\frac{(\omega-\omega_0)^2 \Delta  t^2}{2}} \right) d\omega \nonumber \\
&=&\!\!\! \frac{i\Delta  t}{2(2\pi)^{3/2}} \int e^{i\vec{k}_\perp \cdot \vec{r}_\perp}  \tilde{{\cal E}}_\sigma^{(0)} (\vec{k}_\perp) d\vec{k}_\perp
\Bigl( \int_{-\infty}^{+\infty} e^{i \omega t} e^{-\frac{(\omega+\omega_0)^2 \Delta  t^2}{2}} d\omega  - \int_{-\infty}^{+\infty} e^{i \omega t} e^{-\frac{(\omega-\omega_0)^2 \Delta  t^2}{2}} d\omega  \Bigr) \nonumber \\
&=&\!\!\! \frac{i\Delta  t}{2(2\pi)^{3/2}} \int e^{i\vec{k}_\perp \cdot \vec{r}_\perp}  \tilde{{\cal E}}_\sigma^{(0)} (\vec{k}_\perp) d\vec{k}_\perp
\Bigl( - \int_{+\infty}^{-\infty} e^{-i \omega t} e^{-\frac{(\omega-\omega_0)^2 \Delta  t^2}{2}} d\omega  - \int_{-\infty}^{+\infty} e^{i \omega t} e^{-\frac{(\omega-\omega_0)^2 \Delta  t^2}{2}} d\omega  \Bigr) \nonumber \\
&=&\!\!\! \frac{i\Delta  t}{2(2\pi)^{3/2}} \int e^{i\vec{k}_\perp \cdot \vec{r}_\perp}  \tilde{{\cal E}}_\sigma^{(0)} (\vec{k}_\perp) d\vec{k}_\perp
\Bigl( \int_{-\infty}^{+\infty} e^{-i \omega t} e^{-\frac{(\omega-\omega_0)^2 \Delta  t^2}{2}} d\omega  - \int_{-\infty}^{+\infty} e^{i \omega t} e^{-\frac{(\omega-\omega_0)^2 \Delta  t^2}{2}} d\omega  \Bigr) \nonumber \\
&=&\!\!\! \frac{i\Delta  t}{2(2\pi)^{3/2}} \int e^{i\vec{k}_\perp \cdot \vec{r}_\perp}  \tilde{{\cal E}}_\sigma^{(0)} (\vec{k}_\perp) d\vec{k}_\perp
\int_{-\infty}^{+\infty} \bigl( e^{-i \omega t} - e^{i \omega t} \bigr) e^{-\frac{(\omega-\omega_0)^2 \Delta  t^2}{2}} d\omega   ,
\label{CEstrenF}
\end{eqnarray}
where in going from the second to the third line, we have replaced the variable $\omega$ with $-\omega$ for the first integral in brackets, and in going from the third to the fourth line, we have used $-\int_{+\infty}^{-\infty} f(x) dx = \int_{-\infty}^{+\infty} f(x) dx$.
In the above
\begin{eqnarray}
\tilde{{\cal E}}_\sigma^{(0)} (\vec{k}_\perp) = \frac{1}{2\pi} \int_{-\infty}^{+\infty} e^{-i\vec{k}_\perp \cdot \vec{r}_\perp}  {\cal E}_\sigma^{(0)}(\vec{r}_\perp)  d \vec{r}_\perp .
\end{eqnarray}

To define the electric field strength for
all $z$,
we only need to replace $\omega t$ in Eq.~(\ref{CEstrenF}) by $\omega t - k_z z$ with $k_z = \sqrt{\frac{\omega^2}{c^2}-\vec{k}_\perp^2}$. Thus, the electric field strength
may be written
\begin{eqnarray}
{\cal E}_\sigma(t,\vec{r}_\perp,z)  =
\frac{i\Delta  t}{2(2\pi)^{3/2}} \int e^{i\vec{k}_\perp \cdot \vec{r}_\perp}  \tilde{{\cal E}}_\sigma^{(0)} (\vec{k}_\perp)  \bigl( e^{-i (\omega t - k_z z)} - e^{i (\omega t - k_z z)} \bigr) e^{-\frac{(\omega-\omega_0)^2 \Delta  t^2}{2}} d\vec{k}_\perp  d\omega   .
\label{CEstrenF2}
\end{eqnarray}
Since $\omega_k=ck=c\sqrt{\vec{k}^2_\perp+k_z^2}$ and $\vec{k}_\perp$ has been integrated separately in Eq.~(\ref{CEstrenF2}), we have $\int d\omega = \int \frac{c^2 k_z}{\omega_k} dk_z$. Applying this to Eq.~(\ref{CEstrenF2}) yields
\begin{eqnarray}
{\cal E}_\sigma(t,\vec{r}) &=& \frac{i\Delta  t}{2(2\pi)^{3/2}} \int e^{i\vec{k}_\perp \cdot \vec{r}_\perp}  \tilde{{\cal E}}_\sigma^{(0)} (\vec{k}_\perp) \bigl( e^{-i (\omega_k t - k_z z)} - e^{i (\omega_k t - k_z z)} \bigr) e^{-\frac{(\omega-\omega_0)^2 \Delta  t^2}{2}} \frac{c^2 k_z}{\omega_k} d\vec{k}  \nonumber \\
&=& \frac{\Delta  t}{(2\pi)^{3/2}} \int \Bigl | \tilde{{\cal E}}_\sigma^{(0)} (\vec{k}_\perp) \Bigr| \frac{c^2 k_z}{\omega_k}  e^{-\frac{(\omega-\omega_0)^2 \Delta  t^2}{2}}  \sin\left(\omega_k t - \vec{k} \cdot \vec{r} - \varphi(\vec{k}_\perp)\right) d\vec{k} ,
\label{CEstrenF3}
\end{eqnarray}
where $\varphi(\vec{k}_\perp)$ is the phase function of $\tilde{{\cal E}}_\sigma^{(0)} (\vec{k}_\perp)$.

\end{widetext}

\subsection{Unperturbed group velocity and photon-momentum sums}

By comparing the quantum expression and the classical expression of the electric strength of the light pulses Eqs.~(\ref{QelctS}) and (\ref{CEstrenF3}), we see that
\begin{eqnarray}
|\alpha_{\vec{k},\sigma}| = \frac{\pi \Delta  t}{\sqrt{V \hbar \omega_k}} \Bigl | \tilde{{\cal E}}_\sigma^{(0)} (\vec{k}_\perp) \Bigr| \frac{c^2k_z}{\omega_k}  e^{-\frac{(\omega-\omega_0)^2 \Delta  t^2}{2}} ,
\label{scalarM}
\end{eqnarray}
where we have assumed that $\varphi(\vec{k}_\perp)$ in Eq.~(\ref{CEstrenF3}) equals $\varphi_{\vec{k},\sigma}$  in Eq.~(\ref{QelctS}).

Inserting Eq.~(\ref{scalarM}) into the quantum description of the energy and momentum of the light pulses Eqs.~(\ref{Qenergy}-\ref{QmomentumCube1}) yields
\begin{eqnarray}
\langle E \rangle &=& \sum_{\sigma} \frac{\Delta  t^2}{8 \pi } \int \Bigl | \tilde{{\cal E}}_\sigma^{(0)} (\vec{k}_\perp) \Bigr|^2 \Bigl( \frac{c^2k_z}{\omega_k} \Bigr)^2  e^{-(\omega-\omega_0)^2 \Delta  t^2} \! d \vec{k} \nonumber \\
&=& \frac{\Delta t^2 c^2}{8 \pi } \!\! \int \Bigl | \tilde{\cal E}^{(0)} (\vec{k}_\perp) \Bigr|^2  \Bigl( \frac{ck_z}{\omega_k} \Bigr)^2  e^{-(\omega-\omega_0)^2 \Delta  t^2} \!  d \vec{k} ,
\label{QCenergy}
\end{eqnarray}
\begin{eqnarray}
\langle \vec{p} \rangle \!\!\!&=&\!\!\! \sum_{\sigma} \frac{\Delta  t^2}{8 \pi } \!\! \int \Bigl | \tilde{{\cal E}}_\sigma^{(0)} (\vec{k}_\perp) \Bigr|^2 \frac{\vec{k}}{\omega_k} \Bigl( \frac{c^2k_z}{\omega_k} \Bigr)^2  e^{-(\omega-\omega_0)^2 \Delta  t^2} \!\! d \vec{k} \nonumber \\
\!\!\!&=&\!\!\! \frac{\Delta t^2 c^2}{8 \pi }  \!\! \int \! \Bigl | \tilde{\cal E}^{(0)} (\vec{k}_\perp) \Bigr|^2 \!\! \frac{\vec{k}}{\omega_k}  \Bigl( \frac{ck_z}{\omega_k} \Bigr)^2 \! e^{-(\omega-\omega_0)^2 \Delta  t^2} \!\! d \vec{k} ,
\label{QCmomentum}
\end{eqnarray}
\begin{eqnarray}
M_2  \!\!\!&=&\!\!\! \sum_{\sigma} \frac{\Delta  t^2}{8 \pi } \!\! \int \Bigl | \tilde{{\cal E}}_\sigma^{(0)} (\vec{k}_\perp) \Bigr|^2 \frac{\hbar \omega_k}{c^2} \Bigl( \frac{c^2k_z}{\omega_k} \Bigr)^2  e^{-(\omega-\omega_0)^2 \Delta  t^2} \!\! d \vec{k} \nonumber \\
\!\!\!&=&\!\!\! \frac{\Delta t^2}{8 \pi }  \!\! \int \! \Bigl | \tilde{\cal E}^{(0)} (\vec{k}_\perp) \Bigr|^2 \!\! \hbar \omega_k  \Bigl( \frac{ck_z}{\omega_k} \Bigr)^2 \! e^{-(\omega-\omega_0)^2 \Delta  t^2} \!\! d \vec{k} ,
\label{QmomentumSquare2}
\end{eqnarray}
and
\begin{eqnarray}
M_3 \!\!\!&=&\!\!\! \sum_{\sigma} \frac{\Delta  t^2}{8 \pi } \!\! \int \Bigl | \tilde{{\cal E}}_\sigma^{(0)} (\vec{k}_\perp) \Bigr|^2 \frac{\hbar^2 \omega_k^2}{c^3} \Bigl( \frac{c^2k_z}{\omega_k} \Bigr)^2  e^{-(\omega-\omega_0)^2 \Delta  t^2} \!\! d \vec{k} \nonumber \\
\!\!\!&=&\!\!\! \frac{\Delta t^2}{8 \pi }  \!\! \int \! \Bigl | \tilde{\cal E}^{(0)} (\vec{k}_\perp) \Bigr|^2 \frac{\hbar^2 \omega_k^2}{c}  \Bigl( \frac{ck_z}{\omega_k} \Bigr)^2 \! e^{-(\omega-\omega_0)^2 \Delta  t^2} \!\! d \vec{k} .
\label{QmomentumCube2}
\end{eqnarray}
where we combined the sum over the squared field amplitudes for all polarizations $\sigma$ into the total mode intensity.

Inserting the above results in the relativistic
energy-momentum relation (\ref{standardDis}) with expectation values
of energy and momentum,
we get
\begin{eqnarray}
\mu^2 c^4 = \langle E \rangle^2 - c^2 \langle \vec{p} \rangle^2 = (\langle E \rangle + c \langle \vec{p} \rangle ) ( \langle E \rangle - c \langle \vec{p} \rangle ) .
\label{EnQL}
\end{eqnarray}
If we assume the electric field strength in the $z=0$ plane to also obey a Gaussian distribution, then we have
\begin{eqnarray}
{\cal E}^{(0)} (\vec{r}_\perp) = {\cal E}_0 e^{-\frac{\vec{r}^2_\perp}{2 w^2 }}  \;,\;  \tilde{\cal E}^{(0)} (\vec{k}_\perp) =  {\cal E}_0 w^2 e^{-\frac{\vec{k}^2_\perp w^2}{2}} ,
\label{xy-Gau}
\end{eqnarray}
where ${\cal E}_0$ is the electric field strength at the center of the plane $z=0$, $w$ represents the pulse width (waist) at $z=0$.

Before resuming our calculations, let us review some results about the integral of the function $e^{-s(x^2+y^2)}$:
\begin{eqnarray}
\int_{-\infty}^{+\infty} \int_{-\infty}^{+\infty} e^{-s(x^2+y^2)} dxdy &=&  \int_{0}^{2\pi} \int_{0}^{\infty} e^{-s r^2} rd\theta dr \nonumber \\
&=& \pi \int_{0}^{\infty} e^{-s a} da \nonumber \\
&=& \frac{\pi}{s} ,
\label{square}
\end{eqnarray}
where $a=r^2$. Taking the derivative of Eq.~(\ref{square}) with respect to $s$ yields
\begin{eqnarray}
\int_{-\infty}^{+\infty} \int_{-\infty}^{+\infty} e^{-s(x^2+y^2)} (x^2+y^2) dxdy = \frac{\pi}{s^2} .
\label{square2}
\end{eqnarray}

Inserting Eq.~(\ref{xy-Gau}) into Eq.~(\ref{QCmomentum}) yields $\langle p_x \rangle = \langle p_y \rangle = 0$ because the integrand is an odd function and the integral space is spherically symmetric. This is also easy to see because the light pulses are symmetric in the plane orthogonal with the z-axis.

If we further assume the wavelength is much smaller than the width and
length of the light pulses, we may have $w,c\Delta  t \gg \lambda =
\frac{2\pi c}{\omega_0}$, which implies that the $\vec k_\perp$ is very tiny, which in turn gives
\begin{eqnarray}
e^{-(\omega-\omega_0)^2 \Delta  t^2} \approx  e^{-(k_z-\frac{\omega_0}{c})^2 c^2\Delta  t^2} .
\end{eqnarray}
Using this approximation, we have
\begin{widetext}
\begin{eqnarray}
\langle E \rangle + c \langle \vec{p_z} \rangle
&=& \frac{c^2 \Delta  t^2 {\cal E}^2_0  w^4}{8 \pi } \int  e^{-(k_z-\frac{\omega_0}{c})^2 c^2\Delta  t^2} e^{-\vec{k}^2_\perp w^2} \Bigl(\frac{ck_z}{\omega_k}\Bigr)^{\!2} \Bigl(1+\frac{ck_z}{\omega_k}\Bigr) d \vec{k} \nonumber \\
&\approx& \frac{c^2 \Delta  t^2 {\cal E}^2_0  w^4}{8 \pi } \int  e^{-(k_z-\frac{\omega_0}{c})^2 c^2\Delta  t^2} e^{-\vec{k}^2_\perp w^2} \Bigl( 2 - \frac{5\vec{k}_\perp^2}{2k_z^2} \Bigr) d \vec{k} \nonumber \\
&=& \frac{c^2 \Delta  t^2 {\cal E}^2_0  w^4}{8 \pi } \int_{-\infty}^{\infty}\!\!\int_{0}^{\infty}\!\!\int_{0}^{2\pi}  e^{-(k_z-\frac{\omega_0}{c})^2 c^2\Delta  t^2} e^{-\vec{k}^2_\perp w^2} \Bigl( 2 - \frac{5k_\perp^2}{2k_z^2} \Bigr) k_{\perp}\,d\phi\,dk_{\perp}\,dk_z \nonumber \\
&=& \frac{c^2 \Delta  t^2 {\cal E}^2_0  w^4}{8 \pi } 2 \pi \int_{-\infty}^{\infty} e^{-(k_z-\frac{\omega_0}{c})^2 c^2\Delta  t^2} \Bigl( \frac{1}{w^2} - \frac{5}{4w^4 k_z^2} \Bigr) dk_z \nonumber \\
&=& \frac{c^2 \Delta  t^2 {\cal E}^2_0  w^2}{4} \int_{-\infty}^{\infty} e^{-(k_z-\frac{\omega_0}{c})^2 c^2\Delta  t^2} \Bigl( 1 - \frac{5}{4k_z^2 w^2} \Bigr) dk_z \nonumber \\
&\approx& \frac{\sqrt{\pi} c \Delta  t w^2 {\cal E}^2_0 }{4} \Bigl(1 - \frac{5\lambda^2}{16\pi^2 w^2}\Bigr) ,
\label{plus}
\end{eqnarray}
where we have used $\omega_k = c \sqrt{k_z^2 + k_\perp^2}$ to obtain the second line, and we have evaluated the integrals over transverse momenta using Eq.~(\ref{square2}) and approximated $1/k_z^2 \approx c^2/\omega_0^2 = \lambda^2/(4\pi^2)$ inside the sharply peaked Gaussian integral to obtain the last line.

The other term $\langle E \rangle - c \langle \vec{p_z} \rangle$ in Eq.~(\ref{EnQL}) may be written as
\begin{eqnarray}
\langle E \rangle - c \langle \vec{p_z} \rangle &=& \frac{c^2 \Delta  t^2 {\cal E}^2_0  w^4}{8 \pi } \int  e^{-(k_z-\frac{\omega_0}{c})^2 c^2\Delta  t^2} e^{-\vec{k}^2_\perp w^2} \Bigl( \frac{ck_z}{\omega_k} \Bigr)^{\!2} \Bigl(1-\frac{ck_z}{\omega_k}\Bigr) d \vec{k}  \nonumber \\
&\approx& \frac{c^2 \Delta  t^2 {\cal E}^2_0  w^4}{8 \pi } \int  e^{-(k_z-\frac{\omega_0}{c})^2 c^2\Delta  t^2} e^{-\vec{k}^2_\perp w^2} \frac{\vec{k}_\perp^2}{2 k_z^2}  d \vec{k}  \nonumber \\
&=& \frac{c^2 \Delta  t^2 {\cal E}^2_0  w^4}{16 \pi } \int_{-\infty}^{\infty}\!\!\int_{0}^{\infty}\!\!\int_{0}^{2\pi}  e^{-(k_z-\frac{\omega_0}{c})^2 c^2\Delta  t^2} e^{-\vec{k}^2_\perp w^2} \frac{\vec{k}_\perp^2}{k_z^2} k_{\perp}\,d\phi\,dk_{\perp}\,dk_z \nonumber \\
&=& \frac{c^2 \Delta  t^2 {\cal E}^2_0  w^4}{16 \pi } 2\pi \frac{1}{2 w^4}\int_{-\infty}^{\infty}  e^{-(k_z-\frac{\omega_0}{c})^2 c^2\Delta  t^2} \frac{1}{k_z^2} dk_z \nonumber \\
&=& \frac{c^2 \Delta  t^2 {\cal E}^2_0 }{16} \int_{-\infty}^{\infty}  e^{-(k_z-\frac{\omega_0}{c})^2 c^2\Delta  t^2} \frac{1}{k_z^2}  d k_z ,
\label{minus0}
\end{eqnarray}
where the binomial expansion $\frac{ck_z}{\omega_k} \approx 1 - \frac{k_\perp^2}{2k_z^2}$ has been used to obtain the second line, and the rule of Gaussian-type integral has been used in going from the third to the fourth line.
\end{widetext}

For short laser pulse, we have $k_z \approx \omega_0/c$, and Eq.~(\ref{minus0}) may be calculated as
\begin{eqnarray}
\langle E \rangle - c \langle \vec{p_z} \rangle
&\approx& \frac{c^4 \Delta  t^2 {\cal E}^2_0 }{16 \omega_0^2} \int  e^{-(k_z-\frac{\omega_0}{c})^2 c^2\Delta  t^2} d k_z  \nonumber \\
&=& \frac{c^4 \Delta  t^2 {\cal E}^2_0 }{16 \omega_0^2}  \frac{\sqrt{\pi}}{c\Delta  t} = \frac{\sqrt{\pi} c^3 \Delta  t {\cal E}^2_0 }{16 \omega_0^2} ,
\label{minus}
\end{eqnarray}
where we have used Eq.~(\ref{square2}) in moving from the first to the second line.

Inserting Eqs.~(\ref{plus}) and (\ref{minus}) into Eq.~(\ref{EnQL}) yields
\begin{eqnarray}
\mu=\frac{\sqrt{\pi}\Delta  t w {\cal E}^2_0 }{8\omega_0} \approx \frac{\langle E \rangle}{\omega_0 w c} = \frac{\langle E \rangle}{c^2} \frac{\lambda}{2\pi w}
,
\label{effemas}
\end{eqnarray}
where we have used $\lambda = \frac{2\pi c}{\omega_0}$.

To calculate the group velocity of the above laser pulse, we would need to
first figure out $\langle E \rangle$ and $\langle p_z  \rangle$ through the different combinations of Eqs.~(\ref{plus}) and (\ref{minus}):
\begin{eqnarray}
\langle E \rangle &=& \frac{1}{2} \bigl( \text{Eq.~(\ref{plus})} + \text{Eq.~(\ref{minus})} \bigr) \nonumber \\
&=& \frac{1}{2} \Bigl[ \frac{\sqrt{\pi} c \Delta  t w^2 {\cal E}^2_0 }{4} \Bigl(1 - \frac{5\lambda^2}{16\pi^2 w^2}\Bigr) + \frac{\sqrt{\pi} c^3 \Delta  t {\cal E}^2_0 }{16 \omega_0^2} \Bigr] \nonumber \\
&=& \frac{\sqrt{\pi} c \Delta  t w^2 {\cal E}^2_0 }{8} \Bigl(1 - \frac{5\lambda^2}{16\pi^2 w^2} + \frac{ c^2 }{4 w^2 \omega_0^2} \Bigr) \nonumber \\
&=& \frac{\sqrt{\pi} c \Delta  t w^2 {\cal E}^2_0 }{8} \Bigl(1 - \frac{5\lambda^2}{16\pi^2 w^2} + \frac{ \lambda^2 }{16 \pi^2 w^2} \Bigr) \nonumber \\
&=& \frac{\sqrt{\pi} c \Delta  t w^2 {\cal E}^2_0 }{8} \Bigl(1 -  \frac{ \lambda^2 }{4 \pi^2 w^2} \Bigr) .
\end{eqnarray}
where we have used $\lambda = \frac{2\pi c}{\omega_0}$ in the fourth line. Similarly
\begin{eqnarray}
c \langle \vec{p_z} \rangle &=& \frac{1}{2} \bigl( \text{Eq.~(\ref{plus})} - \text{Eq.~(\ref{minus})} \bigr)  \nonumber \\
&=& \frac{1}{2} \Bigl[ \frac{\sqrt{\pi} c \Delta  t w^2 {\cal E}^2_0 }{4} \Bigl(1 - \frac{5\lambda^2}{16\pi^2 w^2}\Bigr) - \frac{\sqrt{\pi} c^3 \Delta  t {\cal E}^2_0 }{16 \omega_0^2} \Bigr] \nonumber \\
&=& \frac{\sqrt{\pi} c \Delta  t w^2 {\cal E}^2_0 }{8} \Bigl(1 - \frac{5\lambda^2}{16\pi^2 w^2} - \frac{ \lambda^2 }{16 \pi^2 w^2} \Bigr) \nonumber \\
&=& \frac{\sqrt{\pi} c \Delta  t w^2 {\cal E}^2_0 }{8} \Bigl(1 -  \frac{ 3\lambda^2 }{8 \pi^2 w^2} \Bigr) .
\end{eqnarray}
Inserting this into Eq.~(\ref{Ha2}) yields
\begin{eqnarray}
v_{g,0}= \frac{\langle p_z \rangle c^2}{\langle E \rangle} =  \frac{1 -  \frac{ 3\lambda^2 }{8 \pi^2 w^2} }{1 -  \frac{ \lambda^2 }{4 \pi^2 w^2} } c \approx \bigl(1- \frac{\lambda^2}{8\pi^2 w^2} \bigr) c  .
\label{cohevel4}
\end{eqnarray}

With the above assumptions, we can also simplify Eq.~(\ref{QmomentumSquare2}) into
\begin{eqnarray}
M_2\!\!\!&=&\!\!\! \frac{\Delta t^2}{8 \pi }  \!\! \int \! \mathcal{E}_{0}^{2}\,w^{4} \hbar \omega_k  \Bigl( \frac{ck_z}{\omega_k} \Bigr)^2 \! e^{-(\omega-\omega_0)^2 \Delta  t^2} e^{- w^2 k_{\perp}^2}  d \vec{k} \nonumber \\
\!\!\!&=&\!\!\!  \frac{\Delta t^2 \mathcal{E}_{0}^{2}\,w^{4} \hbar c^2}{8 \pi }  \!\! \int \frac{k_z^2}{\omega_k} \!e^{-(k_z-\frac{\omega_0}{c})^2 c^2\Delta  t^2} e^{- w^2 k_{\perp}^2}  d \vec{k} , \nonumber \\
\label{QmomentumSquare3}
\end{eqnarray}
where we have used
$ e^{-(\omega-\omega_0)^2 \Delta  t^2} \approx  e^{-(k_z-\frac{\omega_0}{c})^2 c^2\Delta  t^2}$.

Since we are interested in the cases that satisfy $w \;\gg\; \lambda \;=\;\frac{2\pi c}{\omega_{0}}$, we have
\begin{eqnarray}
\omega_{k}
=
c\,\sqrt{k_{z}^{2} + k_{\perp}^{2}}
\;\approx\;
c\,\Bigl(k_{z} + \tfrac{k_{\perp}^{2}}{2\,k_{z}}\Bigr),
\end{eqnarray}
and
\begin{eqnarray}
\frac{k_{z}^{2}}{\omega_{k}}
\;\approx\;
\frac{k_{z}^{2}}{\,c\,\sqrt{k_{z}^{2}+k_{\perp}^{2}}\,}
\;\approx\;
\frac{1}{c}\,\Bigl(k_{z} - \tfrac{k_{\perp}^{2}}{2\,k_{z}}\Bigr).
\end{eqnarray}
Therefore, Eq.~(\ref{QmomentumSquare3}) becomes
\begin{eqnarray}
M_2 \!&\approx &\! \frac{\Delta t^2 \mathcal{E}_{0}^{2}\,w^{4} \hbar c}{8 \pi }  \!\! \int  k_{z} e^{-(k_z-\frac{\omega_0}{c})^2 c^2\Delta  t^2} e^{- w^2 k_{\perp}^2}  d \vec{k} , \nonumber \\
\label{QmomentumSquare4}
\end{eqnarray}
where we have assumed that $k_{\perp} \ll k_{z}$.
The integral of Eq.~(\ref{QmomentumSquare4}) equals
\begin{widetext}
\begin{eqnarray}
M_2
&=& \frac{\Delta t^2 \mathcal{E}_{0}^{2}\,w^{4} \hbar c}{8 \pi } 2\pi
\int_{-\infty}^{\infty}\!\!\int_{0}^{\infty}
k_{z} e^{-(k_z-\frac{\omega_0}{c})^2 c^2\Delta  t^2} e^{- w^2 k_{\perp}^2}
k_{\perp}  dk_{\perp}  dk_{z} \nonumber \\
&=&
\frac{\Delta t^2 \,\mathcal{E}_{0}^{2}\,w^{4}\,\hbar\,c}{4}
\frac{1}{2\,w^{2}} \int_{-\infty}^{\infty}
k_{z} e^{-c^{2} \Delta t^{2} (k_{z}-\frac{\omega_0}{c} )^{2}}
dk_{z} \nonumber \\
&=&
\frac{\Delta t^2 \,\mathcal{E}_{0}^{2}\,w^{4}\,\hbar\,c}{4}
\frac{1}{2\,w^{2}} \frac{\omega_0}{c^{2}\,\Delta t}\,\sqrt{\pi} \nonumber \\
&=&
\frac{\sqrt{\pi}  \Delta t \,w^2  \mathcal{E}_{0}^{2} \hbar \omega_0  }{8 c}
,
\label{QmomentumSquare5}
\end{eqnarray}
where we have used Eq.~(\ref{square}) in moving from the second to the third line.
\end{widetext}

Similarly,  with the assumption $w \gg \lambda$, we can simplify Eq.~(\ref{QmomentumCube2}) into
\begin{eqnarray}
M_3 \!\!\!&=&\!\!\! \frac{\Delta t^2}{8 \pi }  \!\! \int \! \mathcal{E}_{0}^{2}\,w^{4} \frac{\hbar^2 \omega_k^2}{c}  \Bigl( \frac{ck_z}{\omega_k} \Bigr)^2 \! e^{-(k_z-\frac{\omega_0}{c})^2 c^2\Delta  t^2} e^{- w^2 k_{\perp}^2}  d \vec{k} \nonumber \\
\!\!\!&=&\!\!\!  \frac{\Delta t^2 \mathcal{E}_{0}^{2}\,w^{4} \hbar^2 c}{8 \pi }  \!\! \int k_z^2 e^{-(k_z-\frac{\omega_0}{c})^2 c^2\Delta  t^2} e^{- w^2 k_{\perp}^2}  d \vec{k} , \nonumber \\
\label{QmomentumCube3}
\end{eqnarray}
where we have used $ e^{-(\omega-\omega_0)^2 \Delta  t^2} \approx  e^{-(k_z-\frac{\omega_0}{c})^2 c^2\Delta  t^2}$. Notice that unlike the calculation for $M_2$, the $\omega_k$ term here cancels out algebraically, leaving no need for further approximations in the integrand prior to integration.

The integral of Eq.~(\ref{QmomentumCube3}) equals
\begin{widetext}
\begin{eqnarray}
M_3
&=& \frac{\Delta t^2 \mathcal{E}_{0}^{2}\,w^{4} \hbar^2 c}{8 \pi } 2\pi
\int_{-\infty}^{\infty}\!\!\int_{0}^{\infty}
k_{z}^2 e^{-(k_z-\frac{\omega_0}{c})^2 c^2\Delta  t^2} e^{- w^2 k_{\perp}^2}
k_{\perp}  dk_{\perp}  dk_{z} \nonumber \\
&=&
\frac{\Delta t^2 \,\mathcal{E}_{0}^{2}\,w^{4}\,\hbar^2\,c}{4}
\frac{1}{2\,w^{2}} \int_{-\infty}^{\infty}
k_{z}^2 e^{-c^{2} \Delta t^{2} (k_{z}-\frac{\omega_0}{c} )^{2}}
dk_{z} \nonumber \\
&=&
\frac{\Delta t^2 \,\mathcal{E}_{0}^{2}\,w^{2}\,\hbar^2\,c}{8} \int_{-\infty}^{\infty} \Bigl( (k_z-\frac{\omega_0}{c})^2 + 2\frac{\omega_0}{c}(k_z-\frac{\omega_0}{c}) + \frac{\omega_0^2}{c^2} \Bigr) e^{-c^{2} \Delta t^{2} (k_{z}-\frac{\omega_0}{c} )^{2}} dk_{z} \nonumber \\
&=&
\frac{\Delta t^2 \,\mathcal{E}_{0}^{2}\,w^{2}\,\hbar^2\,c}{8} \Bigl( \frac{\sqrt{\pi}}{2 c^3 \Delta t^3} + 0 + \frac{\omega_0^2}{c^2}\frac{\sqrt{\pi}}{c\Delta t} \Bigr) \nonumber \\
&\approx&
\frac{\Delta t^2 \,\mathcal{E}_{0}^{2}\,w^{2}\,\hbar^2\,c}{8} \frac{\omega_0^2}{c^2} \frac{\sqrt{\pi}}{c\Delta t} \nonumber \\
&=&
\frac{\sqrt{\pi}  \Delta t \,w^2  \mathcal{E}_{0}^{2} \hbar^2 \omega_0^2  }{8 c^2}
,
\label{QmomentumCube4}
\end{eqnarray}
\end{widetext}
where we have used Eq.~(\ref{square}) in moving from the first to the second line, and we have safely ignored the negligibly small term proportional to $1/\Delta t^3$ in moving from the fourth to the fifth line by assuming $\omega_0 \Delta t \gg 1$ for short laser pulses containing many optical cycles.

\section{GUP correction to the pulse group velocity}
\label{app:influence}

We now derive the macroscopic pulse velocity from the per-mode result Eq.~(\ref{Gvelo1}); no factorization of expectation values is assumed anywhere in what follows. Because the sums $M_n$ are built from the wave vector, the per-mode velocity must be taken in the wave-vector form derived above, not by substituting $p \to \hbar k$ into Eq.~(\ref{Gvelo1}); the two differ at quadratic order. For a massless single mode, $pc^2/E = c$ and the velocity points along $\hat k$, giving
\begin{eqnarray*}
v_z(\vec k) &=& c\,\frac{k_z}{k}\Bigl( 1 - \alpha_0 l_{\rm Pl} k + \beta_0^{(k)} l_{\rm Pl}^2 k^2 - \cdots \Bigr), \\
\beta_0^{(k)} &\equiv& \beta_0 + \tfrac{1}{2}\alpha_0^2 .
\end{eqnarray*}
We take the pulse velocity to be the energy-weighted mean of the per-mode velocities, which is the standard energy-transport definition and is the quantity Fedorov and Vintskevich~\cite{fedorov2017} compute at zeroth order:
\begin{eqnarray*}
V_z \;\equiv\; \frac{\sum_{\vec k,\sigma}\langle a_{\vec{k},\sigma}^\dagger a_{\vec{k},\sigma}\rangle\, \hbar\omega_k\, v_z(\vec k)}{\sum_{\vec k,\sigma}\langle a_{\vec{k},\sigma}^\dagger a_{\vec{k},\sigma}\rangle\, \hbar\omega_k} .
\end{eqnarray*}
Because $\hbar\omega_k = \hbar c k$ exactly for photons, the factor $k$ in the weight cancels the $1/k$ in the per-mode velocity exactly, and the numerator becomes
\begin{eqnarray*}
c^2 \sum_{\vec k,\sigma}\langle a_{\vec{k},\sigma}^\dagger a_{\vec{k},\sigma}\rangle\, \hbar k_z \Bigl( 1 - \alpha_0 l_{\rm Pl} k + \beta_0^{(k)} l_{\rm Pl}^2 k^2 \Bigr)  \\
\simeq\; c^2 \Bigl[ \langle p_z\rangle - \alpha_0 \frac{l_{\rm Pl}}{\hbar} M_2 + \beta_0^{(k)} \frac{l_{\rm Pl}^2}{\hbar^2} M_3 \Bigr] ,
\end{eqnarray*}
where the single approximation used is $k_z \simeq k$, which we already invoke for $w \gg \lambda$. The denominator is $\langle E\rangle$ by definition. The macroscopic group velocity of the light pulses under quantum gravity is therefore
\begin{eqnarray}
\langle v_{g} \rangle \!\! &=& \!\!\frac{c^2}{\langle E \rangle} \left( \langle p_z \rangle - \alpha_0 \frac{ l_\text{Pl} }{\hbar} \, M_2 + \beta_0^{(k)} \Bigl(\frac{ l_\text{Pl}}{\hbar}\Bigr)^{\!2}  M_3 - \cdots \right) \nonumber \\
&=& \!\! \frac{c^2 \langle p_z \rangle}{\langle E \rangle} \left( 1 - \alpha_0 \frac{ l_\text{Pl} }{\hbar} \, \frac{M_2}{\langle p_z \rangle} + \beta_0^{(k)} \Bigl(\frac{ l_\text{Pl}}{\hbar}\Bigr)^{\!2}  \frac{M_3}{\langle p_z \rangle} - \cdots \right) \nonumber \\
&=& \!\! v_{g,0}  \left( 1 - \alpha_0 \frac{ l_\text{Pl} }{\hbar} \frac{ M_2}{\langle p_z \rangle} + \beta_0^{(k)} \Bigl(\frac{ l_\text{Pl}}{\hbar}\Bigr)^{\!2} \frac{ M_3}{\langle p_z \rangle} - \cdots \right) , \nonumber \\
\label{GUPveloPulse}
\end{eqnarray}
where the identification $M_1 \simeq \langle p_z \rangle$ uses the same $k_z \simeq k$ approximation stated above ($w \gg \lambda$).

Applying the derived lowest-order result for the macroscopic momentum $\langle p_z \rangle \approx \frac{\sqrt{\pi} \Delta t w^2 {\cal E}_0^2}{8}$ together with Eqs.~(\ref{QmomentumSquare5}) and (\ref{QmomentumCube4}), the GUP modification ratio terms evaluate to
\begin{eqnarray}
\alpha_0 \frac{ l_\text{Pl} }{\hbar} \frac{ M_2}{\langle p_z \rangle} &\approx& \alpha_0 \frac{ l_\text{Pl} }{\hbar} \frac{ \Bigl( \frac{\sqrt{\pi}  \Delta t \,w^2  \mathcal{E}_{0}^{2} \hbar \omega_0  }{8 c} \Bigr) }{ \frac{\sqrt{\pi} \Delta t w^2 {\cal E}_0^2}{8} } \nonumber \\
&=& \alpha_0 \frac{ l_\text{Pl} }{\hbar} \frac{\hbar \omega_0}{c} = \alpha_0 l_\text{Pl} \frac{\omega_0}{c} , \label{GUPterm1} \\
\beta_0^{(k)} \Bigl(\frac{ l_\text{Pl}}{\hbar}\Bigr)^{\!2} \frac{ M_3}{\langle p_z \rangle} &\approx& \beta_0^{(k)} \frac{ l_\text{Pl}^2}{\hbar^2} \frac{ \Bigl( \frac{\sqrt{\pi}  \Delta t \,w^2  \mathcal{E}_{0}^{2} \hbar^2 \omega_0^2  }{8 c^2} \Bigr) }{ \frac{\sqrt{\pi} \Delta t w^2 {\cal E}_0^2}{8} } \nonumber \\
&=& \beta_0^{(k)} \frac{ l_\text{Pl}^2}{\hbar^2} \frac{\hbar^2 \omega_0^2}{c^2} = \beta_0^{(k)} \frac{l_\text{Pl}^2 \omega_0^2}{c^2} . \label{GUPterm2}
\end{eqnarray}

Inserting Eqs.~(\ref{GUPterm1}) and (\ref{GUPterm2}) into Eq.~(\ref{GUPveloPulse}), and substituting the unperturbed analytical group velocity $v_{g,0} \approx c \bigl( 1 - \frac{\lambda^2}{8\pi^2 w^2} \bigr)$ from Eq.~(\ref{cohevel4}), the effective group velocity of the laser pulses with the quantum gravity corrections explicitly becomes
\begin{eqnarray}
\langle v_{g} \rangle &\approx& v_{g,0} \left( 1 - \alpha_0 l_\text{Pl} \frac{\omega_0}{c} + \beta_0^{(k)} \frac{l_\text{Pl}^2 \omega_0^2}{c^2} \right) \nonumber \\
&\approx& v_{g,0} \left( 1 - 2\pi \alpha_0 \frac{l_\text{Pl}}{\lambda} + 4\pi^2 \beta_0^{(k)} \frac{l_\text{Pl}^2}{\lambda^2} \right) .
\label{GUPveloFinal}
\end{eqnarray}

This confirms that the quantum gravity (GUP) modifications to the velocity of the macroscopic laser pulse map to the single-photon scale $\omega_0$ (or wavelength $\lambda$): the extensive macroscopic energy parameter of the pulse scales homogeneously in the numerator and denominator and drops out. We state the logical status of this result plainly: a per-mode deformation is assumed, and a per-photon answer is verified, so the absence of the soccer-ball problem holds by construction once the per-mode prescription is adopted. The non-trivial content is structural: the ratio $M_2/M_1$ selects the central frequency $\omega_0$ rather than a bandwidth-weighted combination, and no off-diagonal mode correlations enter. At this order the result depends only on the spectral occupations $\langle a^\dagger_{\vec k,\sigma} a_{\vec k,\sigma}\rangle$, so any observable distinguishing quantum light from a bright classical pulse must couple to higher photon-number moments, e.g.\ $\mathrm{Var}(\hat n_{\vec k,\sigma})$.

\section{The linear GUP effect}
\label{app:linear}

\subsection{Two-color arrival-time delay}

Having established that the Generalized Uncertainty Principle (GUP) modification to the macroscopic group velocity maps to the single-photon wavelength scale, we propose a ``two-color vacuum chronoscopy'' experiment that exploits the different scaling behaviors of classical diffraction and quantum gravity.

The absolute time-of-flight for a coherent laser pulse propagating over a macroscopic baseline $L$ is given by $t = L/\langle v_g \rangle$. Inserting the inverse of our derived group velocity from Eq.~(\ref{GUPveloFinal}), the expected time-of-flight to leading order is:
\begin{eqnarray}
t(\lambda, w) \approx \frac{L}{c} \left( 1 + \frac{\lambda^2}{8\pi^2 w^2} + 2\pi\alpha_0 \frac{l_\text{Pl}}{\lambda} - 4\pi^2\beta_0^{(k)} \frac{l_\text{Pl}^2}{\lambda^2} \right) . \nonumber \\
\label{app:ToF_Analytical}
\end{eqnarray}

To detect the Planck-scale deformation, we evaluate the relative arrival time difference ($\Delta T$) between a fundamental infrared laser pulse ($\lambda_\text{IR}$) and its corresponding High-Harmonic Generation (HHG) X-ray pulse ($\lambda_\text{X}$). Since the X-ray pulse is generated by the optical field inside a thin noble gas target at the coordinate origin ($z=0$), the two pulses share a common origin, separated by a birth-time offset $\tau_0$ set by the emission window of the harmonic process. We set $\tau_0 = 0$ for the remainder of this appendix, and treat it as a systematic in Appendix~\ref{app:tau0}, where it is removed by a baseline scan rather than assumed away.

After co-propagating through an ultra-high vacuum (UHV) tube of length $L$, the arrival time delay $\Delta T = t(\lambda_\text{IR}, w) - t(\lambda_\text{X}, w)$ evaluates to:
\begin{eqnarray}
\Delta T \approx \frac{L}{c} \Biggl[ \frac{\lambda_\text{IR}^2 - \lambda_\text{X}^2}{8\pi^2 w^2}  \quad + 2\pi\alpha_0 l_\text{Pl} \left( \frac{1}{\lambda_\text{IR}} - \frac{1}{\lambda_\text{X}} \right) \Biggr] ,
\label{app:DeltaT_exact}
\end{eqnarray}
where we have truncated the quadratic $\beta_0^{(k)}$ parameter since it scales with $l_\text{Pl}^2/\lambda^2$ and is expected to be sub-dominant for linearly modified GUP models. Using the exact algebraic identity $\frac{1}{\lambda_\text{IR}} - \frac{1}{\lambda_\text{X}} = -\frac{\lambda_\text{IR}-\lambda_\text{X}}{\lambda_\text{IR}\lambda_\text{X}}$ (no approximation is involved in this step), Eq.~(\ref{app:DeltaT_exact}) becomes
\begin{eqnarray}
\Delta T \approx \frac{L}{c} \Biggl[ \frac{\lambda_\text{IR}^2 - \lambda_\text{X}^2}{8\pi^2 w^2}  \quad - 2\pi\alpha_0 l_\text{Pl} \left( \frac{\lambda_\text{IR}-\lambda_\text{X}}{\lambda_\text{IR}\lambda_\text{X}} \right) \Biggr] .
\label{app:DeltaT_approx}
\end{eqnarray}
Experimentally, the classical term can be disentangled from the GUP term by systematically varying the macroscopic beam waist $w$ via active beam-forming optics at the source. By plotting the measured temporal delay $\Delta T$ against the spatial parameter $1/w^2$, the system will exhibit a linear relationship. The $y$-intercept of this plot (corresponding to the theoretical limit of an infinite plane wave, $1/w^2 \rightarrow 0$) removes the waist-dependent classical diffraction term. Any waist-independent systematic, however, survives at the intercept; two such terms, residual-gas dispersion and the harmonic birth-time offset, are quantified in Appendices~\ref{app:gas} and~\ref{app:tau0} and must be removed independently. After that subtraction, any statistically significant, waist-independent negative delay at this intercept provides a direct measurement of the linear GUP parameter $\alpha_0$.

\subsection{Comparison with astrophysical bounds}

To critically evaluate the practical feasibility of this proposed method, we compute the relevant bounds. Astrophysical measurements already test this exact same free-space photon dispersion relation (kinematic Lorentz Invariance Violation). Space telescopes tracking GeV gamma rays from Gamma-Ray Bursts (e.g., the Fermi-LAT observation of GRB 090510) across billions of lightyears amplify the GUP delay to measurable macroscopic times.

For a cosmological source, the energy-dependent time delay $\Delta t$ resulting from a linear ($n=1$) modification to the group velocity is integrated over the expansion of the universe:
\begin{eqnarray}
\Delta t \approx \alpha_0 \frac{E}{E_\text{Pl}} \frac{1}{H_0} \int_0^z \frac{1+z'}{\sqrt{\Omega_M(1+z')^3 + \Omega_\Lambda}} dz' ,
\end{eqnarray}
where $E$ is the detected photon energy, $E_\text{Pl} = \sqrt{\hbar c^5 / G} \approx 1.22 \times 10^{19}\text{ GeV}$ is the Planck energy, and $H_0 \approx 67.4\text{ km s}^{-1}\text{ Mpc}^{-1}$ is the Hubble constant.

For the short gamma-ray burst GRB 090510, located at a redshift of $z = 0.903$, the Fermi Large Area Telescope (Fermi-LAT) detected a $31.2\text{ GeV}$ photon arriving within a maximum window of $\Delta t \approx 0.829\text{ s}$ from the low-energy prompt emission. Adopting standard $\Lambda\text{CDM}$ cosmological parameters ($\Omega_M \approx 0.315$, $\Omega_\Lambda \approx 0.685$), the effective cosmological time-of-flight evaluates to $\frac{1}{H_0}\int_0^{0.903} \dots dz' \approx 4.58 \times 10^{17}\text{ s}$.

Substituting these observational values into the delay equation yields:
\begin{eqnarray}
0.829\text{ s} &\gtrsim& \alpha_0 \left( \frac{31.2\text{ GeV}}{1.22 \times 10^{19}\text{ GeV}} \right) \left( 4.58 \times 10^{17}\text{ s} \right) \nonumber \\
&\approx& \alpha_0 \left( 1.17\text{ s} \right) .
\end{eqnarray}
This sets a naive observational upper limit of $\alpha_0 \lesssim 0.71$, derived from the raw arrival window of a single photon. The dedicated statistical analyses of the same burst are considerably stronger. Our velocity model is $v(E) = c(1 - \alpha_0 E/E_{\rm Pl} + \cdots)$, expressed in the photon energy $E = cp$, whereas the published limits are stated through the dispersion $E^2 = p^2c^2[1 - s(E/E_{{\rm QG},n})^n]$ with group velocity $v = c[1 - s\,\frac{n+1}{2}(E/E_{{\rm QG},n})^n]$ and $s = +1$ ($-1$) for the subluminal (superluminal) branch. Our $\alpha_0 > 0$ term is subluminal, and the translation at $n=1$ carries unit convention factor, $\alpha_0 = E_{\rm Pl}/E_{{\rm QG},1}$. Reference~\cite{vasileiou2013} obtains $E_{{\rm QG},1} > 9.3\times10^{19}$~GeV ($=7.6\,E_{\rm Pl}$) at 95\% CL in the subluminal linear case, giving $\alpha_0 \lesssim 0.13$; under the maximally conservative treatment of source-intrinsic spectral evolution in the same reference ($E_{{\rm QG},1} > 2\,E_{\rm Pl}$) this degrades only to $\alpha_0 \lesssim 0.5$. The LHAASO observations of GRB~221009A~\cite{lhaaso2024} quote $E_{{\rm QG},1} > 10\,E_{\rm Pl}$ without separating the branches, i.e.\ $\alpha_0 \lesssim 0.10$; the branch-resolved analysis of the same dataset in Ref.~\cite{yang2024} gives $E_{{\rm QG},1} > 1.47\times10^{20}$~GeV in the subluminal case, i.e.\ $\alpha_0 \lesssim 0.083$. Independent analyses of that dataset span roughly a factor of two, and the strongest of them are obtained by fitting an assumed intrinsic light curve and spectral evolution at the source, which is the modeling a terrestrial measurement is designed to avoid. Therefore, any laboratory-based vacuum chronoscopy experiment aiming to isolate and discover linear Planck-scale deformations must possess the sensitivity to resolve $\alpha_0 < 1$ across terrestrial baselines.

\section{The quadratic GUP effect}
\label{app:quadratic}

\subsection{Two-color arrival-time delay}

While the linear Generalized Uncertainty Principle parameter $\alpha_0$ represents a parity-violating correction, many prominent quantum gravity models (such as those motivated by string theory and certain formulations of doubly special relativity) preserve spatial parity and predict a purely quadratic minimal-length modification. In these frameworks, the linear parameter vanishes ($\alpha_0 = 0$), and the leading-order correction to spacetime geometry is governed by the quadratic parameter $\beta_0$. We therefore evaluate the resolving power of the proposed two-color vacuum chronoscopy experiment for the $\beta_0$ parameter separately.

Assuming $\alpha_0 = 0$, the wave-vector and physical-momentum quadratic coefficients coincide, $\beta_0^{(k)} = \beta_0$, so throughout this appendix the distinction drawn after Eq.~(\ref{Gvelo1}) is immaterial and we write $\beta_0$ throughout. The macroscopic group velocity derived from the multimode coherent state expectation in Eq.~(\ref{GUPveloFinal}) then simplifies to
\begin{eqnarray}
\langle v_g \rangle \approx c \left( 1 - \frac{\lambda^2}{8\pi^2 w^2} \right) \left( 1 + 4\pi^2 \beta_0 \frac{l_\text{Pl}^2}{\lambda^2} \right) .
\end{eqnarray}
For $\beta_0 > 0$, the quadratic GUP correction acts to \textit{increase} the group velocity of the photon at high energies (shorter wavelengths). This represents a slight kinematic superluminal advance that opposes the subluminal classical geometric delay.

Inverting this expression to find the expected time-of-flight $t = L/\langle v_g \rangle$ over a macroscopic baseline $L$, and expanding to leading order, yields:
\begin{eqnarray}
t(\lambda, w) \approx \frac{L}{c} \left( 1 + \frac{\lambda^2}{8\pi^2 w^2} - 4\pi^2\beta_0 \frac{l_\text{Pl}^2}{\lambda^2} \right) .
\end{eqnarray}

Following the same differential chronoscopic methodology, the arrival time difference $\Delta T = t(\lambda_\text{IR}, w) - t(\lambda_\text{X}, w)$ between the fundamental infrared pulse and the synchronized attosecond X-ray pulse evaluates to
\begin{eqnarray}
\Delta T \approx \frac{L}{c} \Biggl[ \frac{\lambda_\text{IR}^2 - \lambda_\text{X}^2}{8\pi^2 w^2} - 4\pi^2 \beta_0 l_\text{Pl}^2 \left( \frac{1}{\lambda_\text{IR}^2} - \frac{1}{\lambda_\text{X}^2} \right) \Biggr] .
\end{eqnarray}
Using the exact algebraic identity $\frac{1}{\lambda_\text{IR}^2} - \frac{1}{\lambda_\text{X}^2} = -\frac{\lambda_\text{IR}^2 - \lambda_\text{X}^2}{\lambda_\text{IR}^2 \lambda_\text{X}^2}$ (again no approximation is involved; the observation that $\lambda_\text{IR} \gg \lambda_\text{X}$ enters only later, when the combined factor is evaluated and found to be dominated by the X-ray wavelength), this expression becomes
\begin{eqnarray}
\Delta T \approx \frac{L}{c} \Biggl[ \frac{\lambda_\text{IR}^2 - \lambda_\text{X}^2}{8\pi^2 w^2} + 4\pi^2 \beta_0 l_\text{Pl}^2 \left( \frac{\lambda_\text{IR}^2 - \lambda_\text{X}^2}{\lambda_\text{IR}^2 \lambda_\text{X}^2}  \right) \Biggr] .
\label{app:DeltaT_beta}
\end{eqnarray}

\subsection{Comparison with astrophysical bounds}

We apply the same astrophysical baseline from GRB 090510 to numerically evaluate the constraint on the quadratic GUP parameter $\beta_0$. Assuming a parity-preserving model where the linear term vanishes ($\alpha_0 = 0$), the temporal dispersion (or advance) of the $E \approx 31.2\text{ GeV}$ photon over cosmological distances is dictated by the modified redshift scaling:
\begin{eqnarray}
|\Delta t| \approx \beta_0 \left(\frac{E}{E_\text{Pl}}\right)^{\!2} \frac{1}{H_0} \int_0^z \frac{(1+z')^2}{\sqrt{\Omega_M(1+z')^3 + \Omega_\Lambda}} dz' .
\end{eqnarray}
Evaluating this redshift integral for $z=0.903$ yields an effective quadratic time-of-flight of $\frac{1}{H_0}\int_0^{0.903} \dots dz' \approx 6.68 \times 10^{17}\text{ s}$.

Given the measured maximum arrival time difference of $|\Delta t| \approx 0.829\text{ s}$ for the $31.2\text{ GeV}$ photon, we obtain:
\begin{eqnarray}
0.829\text{ s} &\gtrsim& \beta_0 \left( \frac{31.2\text{ GeV}}{1.22 \times 10^{19}\text{ GeV}} \right)^{\!2} \left( 6.68 \times 10^{17}\text{ s} \right) \nonumber \\
&\approx& \beta_0 \left( 6.54 \times 10^{-36} \right) \left( 6.68 \times 10^{17}\text{ s} \right) \nonumber \\
&\approx& \beta_0 \left( 4.37 \times 10^{-18}\text{ s} \right) .
\end{eqnarray}
This naively bounds the quadratic parameter to $\beta_0 \lesssim 1.9 \times 10^{17}$.

Two qualifications are required. First, since our $\beta_0 > 0$ term is superluminal, the appropriate published comparison is the \emph{superluminal} quadratic branch, which carries the $(n+1)/2 = 3/2$ convention factor in the velocity: $\beta_0 = \frac{3}{2}(E_{\rm Pl}/E_{{\rm QG},2})^2$. The superluminal limit of Ref.~\cite{vasileiou2013}, $E_{{\rm QG},2} > 9.4\times10^{10}$~GeV, then gives $\beta_0 \lesssim 2.5\times10^{16}$. The LHAASO observations of GRB~221009A~\cite{lhaaso2024} quote $E_{{\rm QG},2} > 6\times10^{-8}E_{\rm Pl} = 7.3\times10^{11}$~GeV without separating the branches, i.e.\ $\beta_0 \lesssim 4.2\times10^{14}$, and the branch-resolved analysis of the same dataset in Ref.~\cite{yang2024} gives $E_{{\rm QG},2} > 7.2\times10^{11}$~GeV in the superluminal case, i.e.\ $\beta_0 \lesssim 4.3\times10^{14}$. These are roughly four hundred times stronger than the naive window estimate above. Second, the naive window argument itself bounds a \emph{delay}: for $\beta_0 > 0$ the high-energy photon would arrive \emph{early}, and observing it $0.829$~s after trigger does not bound an advance without an assumption about the latest plausible emission time. This is why the dedicated analyses treat the two branches separately, and why we quote the published superluminal limits rather than the window estimate.

The much weaker astrophysical constraint for the quadratic GUP term stems directly from the extreme $(E/E_\text{Pl})^2 \sim 10^{-36}$ attenuation at GeV energy scales. This implies that testing the quadratic order requires either a large improvement in experimental timing precision over macroscopic baselines, or probing with significantly higher energy sources, to resolve $\beta_0 \sim \mathcal{O}(1)$. For the terrestrial setup itself the quadratic delay over the baseline is $(L/c)\,4\pi^2\beta_0\,l_{\rm Pl}^2/\lambda_{\rm X}^2 \approx 1.4\times10^{-55}\,\beta_0$~s, so a $10^{-21}$~s null yields only $\beta_0 \lesssim 7\times10^{33}$: the proposed method has no meaningful quadratic reach at foreseeable timing precision, and the $\beta_0$ term is discussed in this work solely as a sign discriminator in principle, after subtraction of the residual-gas systematic of Appendix~\ref{app:gas}. (Since this appendix assumes $\alpha_0 = 0$, the terrestrial bound is on $\beta_0^{(k)} = \beta_0$; when both parameters are non-zero it is $\beta_0^{(k)}$ that the wavelength-domain measurement constrains, while the astrophysical limits above constrain $\beta_0$.)

\section{Order-of-magnitude estimates and terrestrial feasibility}
\label{app:estimate}

To evaluate the practical feasibility of the proposed two-color vacuum chronoscopy, we compute the expected temporal delays using current laboratory parameters. This highlights the capacity of our macroscopic coherent setup to establish direct, model-independent terrestrial bounds on the GUP parameters, decoupling the search for quantum gravity from unverified astrophysical emission mechanisms.

\subsection{Experimental parameters}
We assume a controlled ultra-high vacuum baseline comparable to modern gravitational wave observatories, yielding a macroscopic propagation length of $L = 4000\text{ m}$.
For the two-color probe, we utilize a fundamental Ti:Sapphire infrared laser pulse ($\lambda_{\text{IR}} = 800\text{ nm}$) and its synchronized high-harmonic generation (HHG) soft X-ray pulse ($\lambda_{\text{X}} = 1\text{ nm}$). Active beam-forming optics tune the macroscopic transverse beam waist around $w \sim 0.1\text{ m}$. We use standard fundamental constants: $c \approx 3 \times 10^8\text{ m/s}$ and $l_{\text{Pl}} \approx 1.616 \times 10^{-35}\text{ m}$.

\subsection{Calibration of the classical diffractive delay}
To successfully isolate the pure GUP signature at the mathematical infinite-waist intercept ($1/w^2 \to 0$), the experiment must first resolve the waist-dependent classical diffractive delay (the first term in Eq.~(\ref{eq:DeltaT_approx})). Since $\lambda_{\text{IR}} \gg \lambda_{\text{X}}$, the classical delay difference is entirely dominated by the fundamental IR pulse:
\begin{equation}
\Delta T_{\text{diff}} \approx \frac{L}{c} \left( \frac{\lambda_{\text{IR}}^2}{8\pi^2 w^2} \right).
\end{equation}
For a controlled beam waist of $w = 0.1\text{ m}$, the temporal walk-off evaluates to:
\begin{equation}
\Delta T_{\text{diff}} \approx \left( \frac{4000}{3 \times 10^8} \right) \frac{(8 \times 10^{-7})^2}{8\pi^2 (0.1)^2} \approx 1.08 \times 10^{-17}\text{ s}.
\end{equation}
This classical geometric walk-off of $\sim 10.8\text{ attoseconds}$ (as) is within the resolution capabilities of modern attosecond metrology (e.g., attosecond streaking or RABBITT techniques, which routinely achieve precision well below $10\text{ as}$). By actively tuning the waist $w$ (e.g., from $0.1\text{ m}$ to $0.2\text{ m}$), the experiment generates measurable delay variations, allowing a linear regression against $1/w^2$ to extract the infinite-waist intercept.

\subsection{Terrestrial bound on the linear GUP parameter}
At the intercept ($1/w^2 \to 0$), classical diffraction is removed, isolating the waist-independent part of the delay. Using Eq.~(\ref{GUPveloFinal}), the absolute pure GUP delay is
\begin{equation}
|\Delta T_{\text{GUP}}| \approx \frac{L}{c} 2\pi\alpha_0 \left(\frac{l_{\text{Pl}}}{\lambda_{\text{X}}}\right).
\end{equation}
Substituting the physical parameters, the delay per unit of $\alpha_0$ evaluates to
\begin{eqnarray}
|\Delta T_{\text{GUP}}| &\approx& \left(\frac{4000}{3 \times 10^8}\right) \times 2\pi \times \left(\frac{1.616 \times 10^{-35}}{10^{-9}}\right) \times \alpha_0 \nonumber \\
&\approx& 1.35 \times 10^{-30} \, \alpha_0\text{ s}.
\end{eqnarray}

Currently, the absolute frontier of direct time-domain chronoscopy (e.g., charting photoemission delays) has breached the zeptosecond barrier ($\delta t_{\text{res}} \sim 10^{-21}\text{ s}$) (see, e.g., Ref.~[\onlinecite{grundmann2020}]). Assuming a measurement limit of $\sim 10^{-21}\text{ s}$ via extensive photon integration and statistical averaging, a null observation at the intercept, after subtraction of the residual-gas term derived in Appendix~\ref{app:gas}, would impose a terrestrial kinematic bound of
\begin{equation}
\alpha_0 \lesssim \frac{10^{-21}\text{ s}}{1.35 \times 10^{-30}\text{ s}} \approx 7.4 \times 10^8.
\end{equation}
While numerically weaker than the published cosmological limits ($\alpha_0 \lesssim 0.08$--$0.13$~\cite{vasileiou2013,lhaaso2024,yang2024}, or $\alpha_0 \lesssim 0.5$ under maximally conservative source assumptions), this represents a controlled, repeatable, source-independent terrestrial measurement of Planck-scale geometry. It circumvents the systemic uncertainties affecting astrophysical observations, such as unknown intergalactic dispersion properties and the unverified assumption of simultaneous multi-wavelength emission at the transient cosmic source; as the conservative limit above shows, those uncertainties cost the cosmological bound a factor of a few, not orders of magnitude, and we make no claim that the terrestrial setup is currently competitive at the Planck scale.

\section{Waist-independent systematic I. Residual-gas dispersion}
\label{app:gas}

The intercept method removes only \emph{waist-dependent} terms, so any waist-independent contribution to $\Delta T$ survives at the intercept and must be budgeted explicitly. Two such terms arise. The first, treated here, is the group delay of the infrared pulse in the residual gas of the vacuum system.

The pulse delay over the baseline is $(L/c)(n_g - 1)$, with $n_g - 1 = (n-1) - \lambda\,dn/d\lambda$ the group refractivity. From the Edl\'en-type dispersion formula for standard air at $\lambda_{\rm IR} = 800$~nm, $(n-1) = 2.750\times10^{-4}$ and $(n_g - 1) = 2.800\times10^{-4}$ (a 1.8\% distinction, negligible here). Scaling to $P = 10^{-9}$~torr $= 1.32\times10^{-12}$~atm gives, for air-like composition, $(n_g-1) = 3.7\times10^{-16}$ and a delay of $4.9\times10^{-21}$~s over $L = 4$~km. Ultra-high-vacuum residual gas at this pressure is in practice hydrogen-dominated. The two refractivities must be compared per molecule rather than at their respective standard states: the Edl\'en value refers to $15\,^\circ$C, where $N_{\rm air} = 2.547\times10^{25}$~m$^{-3}$, whereas $(n-1)_{\rm H_2} \approx 1.38\times10^{-4}$ refers to $0\,^\circ$C, where $N_{\rm H_2} = 2.687\times10^{25}$~m$^{-3}$. The per-molecule ratio is then $0.467$ rather than the $0.50$ obtained by dividing the two tabulated refractivities directly, and the hydrogen delay is $2.3\times10^{-21}$~s. On the X-ray side, $n = 1 - \delta$ with $\delta = r_e \lambda_{\rm X}^2 n_e/2\pi \approx 3.0\times10^{-20}$ (H$_2$) at this pressure; since $\delta \propto \omega^{-2}$, the X-ray \emph{group} index is $1 + \delta$, so the X-ray pulse is also delayed, but only by $4\times10^{-25}$~s, four orders of magnitude below the infrared term. The differential delay is therefore infrared-dominated and equals the numbers above.

Three properties of this term matter. (i)~It exceeds the assumed $10^{-21}$~s timing floor by a factor of 2.3 (H$_2$) to 4.9 (air-like); in $\alpha_0$ units it corresponds to $1.7\times10^{9}$ and $3.6\times10^{9}$ respectively, against the bound of $7.4\times10^{8}$ derived above. (ii)~It is independent of $w$, so it sits precisely in the regression intercept. (iii)~The gas delays the infrared pulse relative to the X-ray pulse, contributing \emph{positively} to $\Delta T = t_{\rm IR} - t_{\rm X}$, the same sign as a $\beta_0 > 0$ advance and opposite to the $\alpha_0 > 0$ delay. Left uncharacterized, the gas term therefore mimics a quadratic-GUP signature, partially masks a linear one, invalidates the sign-of-intercept discrimination, and limits the achievable bound to $\alpha_0 \lesssim 1.7\times10^{9}$: the experiment would be systematics-limited rather than timing-limited.

The gas delay is strictly linear in the column density, hence in pressure at fixed composition and temperature, while the GUP term is pressure-independent, so a pressure scan is the natural handle. It must, however, be constructed correctly. Backfilling a test gas and regressing on the added pressure does \emph{not} work: the pre-existing residual sits at fixed partial pressure across every step of such a scan, so it contributes a constant rather than a slope and remains in the intercept, exactly degenerate with the quantity being measured. The fitted slope then reports the refractivity of the backfilled species and subtracts only the gas that was deliberately added. What is required is to vary the \emph{base} pressure at approximately fixed composition, for example by throttling the pumping speed over $10^{-10}$--$10^{-8}$~torr, and to include the measured total pressure as a second regression variable alongside $1/w^2$; the extrapolation to $P \to 0$ then follows the same logic as the waist regression. Any residual change of composition across the scan enters as a second-order correction and should be monitored with a residual-gas analyzer. Two alternatives avoid the scan altogether: residual-gas-analyzer characterization of the column composition together with total-pressure gauging to better than $\sim 40\%$ (H$_2$) suppresses the subtraction residual below $10^{-21}$~s, and a 10\% characterization leaves a systematic floor of $2.3\times10^{-22}$~s, i.e.\ $\alpha_0 \approx 1.7\times10^{8}$, restoring the timing-limited regime; alternatively, operating below $P \lesssim 4\times10^{-10}$~torr of H$_2$ renders the term subdominant outright, and this should be stated as a design requirement for a dedicated installation. All terrestrial bounds quoted in this work assume one of these.

\section{Waist-independent systematic II. Harmonic birth-time offset}
\label{app:tau0}

The two pulses are not born simultaneously. The harmonic burst is emitted within a fraction of the driving optical half-cycle, so the birth-time offset $\tau_0$ between the infrared envelope and the X-ray burst is of order $10^{-16}$ to $10^{-15}$~s at $\lambda_{\rm IR} = 800$~nm, five to six orders of magnitude above the target sensitivity. Because the beam-forming optics act downstream of the gas target, the driving intensity at the target, and hence $\tau_0$, does not change as $w$ is varied. The offset is therefore waist-independent and survives the $1/w^2$ extrapolation, in the same way as the residual-gas term and with a far larger magnitude.

It is nevertheless removable without being known. Writing the measured quantity as
\begin{eqnarray*}
\Delta T(w,L) = \tau_0 + \frac{L}{c}\Bigl[ D(w) + G + \Gamma \Bigr] ,
\end{eqnarray*}
with $D(w)$ the diffractive term, $G$ the residual-gas term and $\Gamma$ the GUP term, the waist regression at fixed $L$ returns an intercept $I(L) = \tau_0 + (L/c)(G + \Gamma)$. Since $\tau_0$ carries no factor of $L$ while $D$, $G$ and $\Gamma$ all do, a second regression of $I(L)$ against $L$ separates the two: the slope isolates $(G+\Gamma)/c$ and the intercept returns $\tau_0$ itself as a by-product. A dedicated installation must therefore provide at least two, and preferably several, propagation lengths; this is an additional design requirement beyond the variable waist, and we state it as such.

Two caveats attach. First, shot-to-shot fluctuation of the driving intensity modulates $\tau_0$ through the intensity-dependent dipole phase of the harmonic emission; this enters as a statistical term and averages down with the photon integration already assumed, but slow drift of the driving conditions does not, and must be bounded over the duration of a baseline scan. Second, the beam-forming optics themselves contribute waist-dependent path terms; these bias the extracted intercept unless they scale as $1/w^2$, which must be verified for a specific optical design rather than assumed. Thermal path-length gradients along the baseline are common-mode between the two colors to leading order and do not contaminate the intercept in the way the residual gas and $\tau_0$ do.

\end{document}